# Radiomics-based artificial intelligence (AI) models in colorectal cancer (CRC) diagnosis, metastasis detection, prognosis, and treatment response

Parsa Karami[1], Reza Elahi[2*]

[1]School of Medicine, Zanjan University of Medical Sciences, Zanjan, Iran

[2]Department of Radiology, Zanjan University of Medical Sciences, Zanjan, Iran

*Corresponding author's email: rezaelahi96research@gmail.com

## Abstract

With a high rate of morbidity and mortality, colorectal cancer (CRC) ranks third in mortality among cancers. By analyzing the texture properties of images and quantifying the heterogeneity of tumors, radiomics and radiogenomics are non-invasive methods to determine the biological properties of CRC. Recently, several articles have discussed the application of radiomics in different aspects of CRC. Therefore, given the large amount of data published, this review aims to discuss how radiomics can be used for distinguishing benign and malignant colorectal lesions, diagnosing, staging, predicting prognosis and treatment response, and predicting lymph node and hepatic metastasis of CRC, based on radiomic features extracted from magnetic resonance imaging (MRI), computed tomography (CT), esophageal ultrasonography (EUS), and positron emission tomography-CT (PET-CT). Challenges in bringing radiomics to clinical application and future solutions have also been discussed. With the progress made in radiomics and the application of deep and machine learning in this area, radiomics can become one of the main tools for the personalized management of CRC patients shortly.

## Introduction

Being the third rank in cancer-related mortality, colorectal cancer (CRC) causes an estimated 826,000 deaths per year (1). The 5-year survival rate of CRC is 11-65%, depending on the stage it is diagnosed (https://www.cancer.org/cancer/colon-rectal-cancer/about/key-statistics.html#:~:text=Excluding%20skin%20cancers%2C%20colorectal%20cancer,new%20cases%20of%20rectal%20cancer) (2). Due to the high heterogeneity of CRC, one of its natural characteristics, the diagnosis, staging, evaluation of treatment response, and prediction of lymph node/hepatic metastasis are major challenges of CRC management (3). Contrast-enhanced computed tomography (CT) is the standard imaging modality for CRC diagnosis and staging *(Figure 1)*. Nevertheless, the modest accuracy of current diagnostic methods necessitates the development of novel strategies with higher accuracy for CRC diagnosing and staging. Therefore, novel artificial intelligence (AI)-based models have been developed to overcome these barriers. Radiomics is a recently developed quantitative imaging method that extracts quantitative radiomic features from medical images (4). The analysis of the quantitative features can enhance the accuracy of cancer diagnosis, staging, response to treatment, lymph node metastasis, and prognosis (5, 6). In this article, we aim to review the radiomics workflow and the progress of AI-assisted radiomic-based models to distinguish malignant and benign colorectal lesions,  as well as diagnosis, staging, predicting prognosis and treatment outcomes, and predicting lymph node and hepatic metastasis of CRC. Moreover, current limitations and potential solutions for achieving higher accuracy are discussed.

## 1. Radiomics workflow *(Figure 2)*

Radiomic-based studies are required to take five steps. The first step is the acquisition of high-quality medical images. Regarding CRC, different imaging methods, such as positron emission tomography (PET), computed tomography (CT), Magnetic resonance imaging (MRI), and Endoscopic ultrasound (EUS) modalities can be used. To have accurate results, medical images should be acquired with high quality. Quantitative imaging biomarker alliance and quantitative imaging network are examples of standard protocols developed for the acquisition of high-quality medical images (https://www.rsna.org/research/quantitative-imaging-biomarkers-alliance, https://imaging.cancer.gov/informatics/qin.htm). The next step is identifying and segmenting the region of interest (ROI), which can be conducted manually, semi-automatically, or automatically. The third step is the extraction of quantitative features from these medical images. Radiomic features are divided into four groups. Shape, texture, border heterogeneity, volume, and surface-to-volume ratio are examples of the first group of radiomic features. The second group of features, also named first-order statistics, are the histogram-based features, such as entropy, standard deviation, kurtosis, skewness, etc. Gray-level size zone matrix (GLSZM) and gray-level co-occurrence matrix (GLCM) are the third group of features, also named second-order statistics, and are used to extract textural features. The fourth group of features, also named higher-order statistics, such as Laplacian of Gaussian and wavelet, is the fourth group. After extracting the features, the fourth step is to analyze the extracted features. Machine learning (ML) and deep learning (DL) methods, such as least absolute shrinkage and selection operator (LASSO), are the most favorite strategies used for feature analysis in radiomics projects. The last step of the radiomic workflow is developing AI-based models using support vector machine (SVM) and linear regression to produce predictive models based on these features (7, 8). In the next section, we overview the current studies of radiomics in evaluating diagnosis and classification, treatment response, lymph node and hepatic metastasis, and prognosis/survival of CRC *(Table 1)*.

## 2. Radiomics in CRC diagnosis

One of the most important issues in CRC is accurately distinguishing malignant tumors from benign ones using both clinical symptoms and paraclinical modalities. The most common imaging modality used to diagnose CRC is CT; however, liver metastases are difficult to detect by this method. In certain cases, tissue biopsies and MRI can be used; however, these methods can be aggressive and may delay the diagnosis of patients and treatment initiation (9). Recent studies consider that adapting AI-based radiomic models to clinical decision-making may improve diagnosing CRC and its metastasis accuracy.

The differentiation of high and low-grade colorectal adenocarcinoma is a clinical challenge since, in most studies, the clinical characteristics of patients with high-grade and low-grade colorectal adenocarcinoma are comparable, except for the location of the tumor-determined by CT (10). Based on quantitative features from contrast-enhanced CT images, radiomics signatures have been developed to preoperatively distinguish the accurate histopathologic differentiation of colorectal adenocarcinoma, which could reduce the need for preoperative invasive diagnostic procedures, such as tissue biopsy. In a study,

Huang et al. developed a radiomic signature to differentiate low and high-grade rectal adenocarcinoma based on CT imaging without the necessity of invasive procedures (AUC=0.812 in the training dataset and 0.735 in the validation dataset). This can also affect the treatment of the patients since neoadjuvant therapy may improve outcomes in high-risk patients identified with the radiomics method. This radiomics signature could also distinguish histologic grade in the colon and rectal adenocarcinomas in stratified analyses (10). In another study by Caruso et al., using CT scan data, radiomic models were developed to distinguish high-risk and low-risk lesions in non-metastatic patients. Their model could appropriately predict high-risk patients with an AUC of 0.73 (11).

A convolutional neural network (CNN) is an AI network that detects colorectal lesions that a manual colonoscopy or capsule endoscopy imaging could miss. A study found that a radiomic-based CNN algorithm had 99.2% sensitivity and 99.6% specificity in distinguishing between the two different types of colon lesions, including mucosal and blood lesions (12). PyRadiomics is a powerful open-source tool for extracting radiomics data from different imaging modalities and has recently gained much attention in CRC radiomics studies. In a study using PyRadiomics for feature extraction, T2-W textural analysis indicated the greatest radiomics performance in classifying liver mucinous metastasis. Furthermore, the best performance was achieved with T2-W extracted features with k-nearest neighbors (KNN) (13).

Detection of perineural invasion (PNI) of colorectal cancer is a critical issue since it impacts the surgical management of colorectal cancer. In a study, Li et al. developed a CT texture-based ML model to predict preoperative PNI status in 207 colon cancer patients. The multi-stage classifier had an AUC=0.793 in detecting PNI in the test setting (14). In another study, Haung et al. developed a CT-based radiomics model in combination with carcinoembryonic antigen (CEA), which showed an appropriate discrimination performance (C-index=0.817) (15).

Colon polyps are tissue growths that could be benign or malignant. Malignant polyps have a high chance of developing malignant CRC (16). It would be helpful if we could distinguish benign from malignant colorectal polyps to identify patients with a high risk of colorectal cancer from those who have a low risk. In a study, although it was not possible to differentiate benign from malignant polyps using CT colonography imaging, machine learning was able to make the distinction based on CT colonography images with adequate sensitivity and specificity of about 82% and 85%, respectively. In this study, the authors determined that the gray-level histogram statistics was the most powerful tool for differentiating the polyps (17). Another study used computer-aided diagnosing (CAD) to classify colorectal polyps found by endoscopic images. This study used AdaBoost ensemble learning, principal component analysis, and a modified deep residual network to detect malignant polyps. This study demonstrated a sensitivity and specificity of 98.82% and 99.38% in automatically detecting benign and malignant colonic polyps (18). Serrated polyps have a higher probability of developing colorectal cancer. By adapting the deep convolutional neural network (DCNN) for reviewing images, CT colonography could detect serrated polyps. In a study, the sensitivity of detecting serrated polyps greater than 6 mm was 93±7%, averaging a 0.8±1.8 false positive rate (19).

## 3. Radiomics in CRC staging and disease progression

Preoperative staging of the CRC significantly affects choosing the proper treatment method for each patient. It has been demonstrated that CT-obtained texture features could identify preoperative colorectal cancer patients in stages I-II and III-IV more accurately than staging only based on tumor-sized measurements. This study showed a sensitivity of 0.611, a specificity of 0.680, and an AUC of 0.792 in staging colorectal cancer. Therefore, texture features could be a reliable method for predicting and staging preoperative colorectal cancer patients (20).

This study aims to determine whether a gene mutation can be detected using radiologic characteristics without invasive procedures such as biopsies or surgery. Using a texture analysis of CT images, Castro et al. demonstrated high accuracy in detecting colorectal cancer related to KRAS mutations (21).

In a study, Wu et al. investigated the potential of radiomics for finding the preoperative MSI status of colorectal cancer patients using dual-energy computed tomography (DECT) imaging modality. They analyzed iodine-based material decomposition (MD) images and concluded that radiomics analysis could properly predict preoperative MSI status with an AUC of 0.918, a sensitivity of 0.875, and a specificity of 0.857 (22). In another study, Li et al. accomplished the radiomics model based on CT images to predict MSI status preoperatively. Their model was efficient in the prediction of CRC patients with MSI (23). Moreover, Fan et al. reported that a combination of CT-radiomics analysis of stage II colorectal cancer with clinical factors, including smoking, age, hypertension, etc, could predict MSI status accurately (24-26).

CT staging, in combination with radiomics, could improve the detection of high-risk CRC. In a study, 292 patients were evaluated with preoperative CT images to identify high-risk cancers. Compared to the CT images only or radiomics features alone, the combination method achieved better performances with an AUC of 0.727 (27).

In a systematic review and meta-analysis on preoperative lymph nodestaging by Bedrikovetski et al., the authors evaluated seventeen studies, including five DL and twelve radiomics models. Studies indicated the higher performance of radiomics and AI-based models in predicting preoperative lymphy node staging compared to the radiologist alone. Regarding rectal cancer, the radiologist achieved an AUROC of 0.676, while the radiomics models achieved a per-patient AUC of=0.808. DL-based models achieved an AUROC=0.917. Furthermore, regarding colorectal cancer, radiomics models showed a per-patient AUROC of 0.727, which was higher than the radiologist alone, who had an AUROC of 0.676. Based on CT/MRI imaging of 10 years, this study showed significant evidence of AI efficacy in predicting lymph node staging and metastasis. (28). In another study, Granata et al. designed a systematic review to investigate radiomics studies between 2010 and 2021 for mCRC patients; ten were included. The study indicated the efficacy and potential of radiomics in predicting liver metastasis recurrence after surgery, lesion response to chemotherapy, and characterization of metastatic lesions (29).

## 4. Radiomics in CRC treatment response

In patients with CRC, it is imperative to predict the outcome of their treatment success with neoadjuvant chemotherapy, chemotherapy, surgery, or biological drugs and determine the next step of their treatment strategy. Radiomic-based AI techniques have consistently been used to improve cancer patient decision-making and disease management. One of these challenges is predicting pathological complete response (CR) after treatment in CRC patients. Accurate prediction of treatment response can aid in choosing the best therapeutic method for each patient and thus improve the treatment success rate and prognosis in CRC patients.

### 4.1. Neoadjuvant chemotherapy (nCRT)

Predicting the complete pathologic response in CRC patients undergoing nCRT is a major clinical challenge since only ¼ of the patients experience a complete pathological response after nCRT. The prediction of complete pathological response has gained more attention since it can prevent unnecessary surgical resection. In a study that included colorectal cancer patients with liver metastases who received their first round of chemotherapy, the effectiveness of their nCRT was predicted using a radiomics-based DL technique. The ResNet10 model was fed with contrast-enhanced multidetector computed tomography (MDCT) images, and the outcome was response prediction. The carcinoembryonic antigen (CEA) was chosen as the predictive clinical factor in this case. With the combination of the CEA level with the DL-based model, an AUC of 0.935 was achieved in the training cohort, and an AUC of 0.830 was achieved in the validation cohort. Therefore, compared to the traditional classifier-based radiomics model (AUC=0.903), combining CEA with the DL model provided more accurate outcomes in predicting neoadjuvant chemotherapy response in liver metastasis of colorectal cancer (30).

Several studies have evaluated the application of radiomics in predicting the outcome of treatment with nCRT in locally advanced rectal cancer patients. Based on a review article published in 2023, 43 articles focused on predicting treatment outcomes with neoadjuvant chemotherapy (31). Some of these articles are summarized in *Table 1*. For example, in a retrospective study in 2022 on 137 patients, the authors applied T2-weighted MRI-derived radiomics features to predict response to treatment with N-Acetyl Cysteine (NAC). Peritumoral and intratumoral feature-based radiomics had an AUC of 0.838 and 0.805, respectively. The combination of intra and peritumoral radiomics features had a higher AUC=0.844. The nomogram combining radiomics and clinical features showed an AUC=0.871, higher than other models (32). In 2020, a systematic review was conducted to evaluate the outcome and survival using radiomic-based features. Of the 76 studies in the systematic review, 10 used 18F-FDG-PET/CT, 30 used CT, and 41 used MRI. Among these studies, only 13 were high-quality, and all had used MRI. This systematic review concluded that MRI-based radiomics features accurately predicted outcomes and survival in LRCM. Moreover, tumors with high heterogeneity had a lower response to treatment, which was attributed to the higher mutations in these tumors (5).

Microsatellite instability (MSI) is the repeated DNA regions caused by the impairment of proteins that contribute to repairing DNA mismatches, named the mismatch repair mechanism. MSI is divided into three types: stability, high-frequency, and Low-frequency MSI. High-frequency MSI has the best prognosis in CRC patients, and studies have recommended that there is no requirement for treatment with adjuvant chemotherapy in these patients (23, 33, 34). Therefore, detecting MSI can significantly affect CRC patients' prognosis and treatment strategy. In a study by Li et al., the authors showed that CT imaging data and clinical features of colorectal cancer can give us some validating data for predicting the MSI occurrence in CRC patients (35). In other studies, Pernicka et al. and Fan et al. showed the effectiveness of radiomics in predicting MSI status in CRC patients (23-25).

### 4.2. Chemotherapy

Chemotherpy-based treatment strategies can be used to reduce the metastases of patients with individual colorectal cancer liver metastases (lmCRC). Nevertheless, it is important to know that only 60% of lmCRC patients will benefit from first-line chemotherapy treatment, and 50% will experience disease progression *(Figure 3)* (36). Radiomics can be used to predict response to first-line chemotherapy regimens in CRC patients. A study used CT-based radiomic features to predict responder and non-responder patients to first-line chemotherapy. The responder's lmCRCs displayed statistically greater diameters than non-responder patients, with a mean size of 25 mm (SD: 14 mm). This study showed that delta radiomics could be a predictable method for evaluating the response of lmCRC to first-line oxaliplatin-based chemotherapy (37).

In a study, baseline CT imaging was used to evaluate the response of liver metastatic lesions to the cytotoxic treatment by FOLFOX/FOLFIRI. Texture analysis indicated a reasonable performance for predicting cytotoxic treatment response of metastatic lesions. An external cohort validated that low skewness predicts a higher response to FOLFOX/FOLFIRI (38).

Response evaluation criteria in solid tumors (RECIST) are developed to predict the response to treating patients with solid malignancies (https://recist.eortc.org/). A study reported that in contrast to RECIST, low homogeneity, and high entropy after chemotherapy helped predict earlier response to the chemotherapy treatment. In this study, a lower IDM, correlation, and ASM and a higher entropy, contrast, and variance were associated with better chemotherapy response (AUC= 0.602-0.784). According to these results, radiomics texture could predict response to chemotherapy with reasonable specificities and sensitivities (39, 40). In another retrospective analysis of 192 patients with colorectal cancer liver metastasis who received first-line chemotherapy, a radionics-based DL model was compared to RECIST criteria in terms of treatment response prediction. The DL model received an AUC of 0.903 in the training and 0.820 in the validation cohort. However, the combination of the DL model with the CEA level showed a slightly improved prediction of chemotherapy response with an AUC of 0.935 in the training and 0.830 in the validation cohort (41).

In a study, radiomic signatures were used to predict the survival outcome of unresectable hepatic metastases in patients undergoing chemotherapy using bevacizumab and FOLFIRI. In this study, a follow-up CT after two months was used as the predictor of poor overall survival of patients. In the validation, external, and training cohorts, results showed a lower OS for patients with a SPECTRA score > 002, which could be reliable for the decision-making of a patient's treatment method. In conclusion, CT texture features could predict first-line chemotherapy in MCRC patients (42). Consistently, results of a study showed that the radiomics-clinical nomogram was capable of better predicting non-responsive and responsive liver metastasis of CRC patients to chemotherapy, based on the MRI-radiomics signature, CA19-9 status, and clinical N staging results (43).

### 4.3.Surgical methods

The findings of another study have shown potential clinical implications since they might prove useful in identifying rectal cancer patients who would benefit from low-risk surgery after chemoradiation. In this study, the authors used texture features of the rectal environment, including lumen or wall, in T2-weighted MRI post-neoadjuvant chemoradiation therapy (nCRT) for predicting tumor regression. They concluded that following nRCT, radiomic properties of the rectal wall were correlated with pathologic tumor stages, regardless of which scanner and institution used (44).

Radiofrequency ablation (RFA) is an invasive procedure used in CRC with lung metastasis and improves CRC patients' survival and life quality (45). Clinical and radiological data with adapting ML could improve the prediction of RFA efficacy in lung metastatic CRC. In a retrospective study, 48 patients were evaluated for lung metastasis before and after RFA admission. In combination with clinical data, adapting the LASSO Cox regression model revealed a beneficial performance of radiomics for diagnosing the most important nodules, which have the highest risks for progression before and after the procedure (46).

### 4.4. Biologic anti-cancer drugs

Biologic drugs are a subgroup of medications used to treat cancer by interfering with cancer metabolism, cell proliferation, or cell growth and have gained much attention in recent years (47). Biological anti-cancer drugs include vascular endothelial growth factor (VEGFR) targeting monoclonal antibodies. Radiomic-based AI networks could also be applied to predict responses to biological drugs. A study proved the efficacy of the DL network method in evaluating early detecting responses to anti-VEGFR therapies in metastatic colon cancer. DL Network detects tumor morphological changes based on CT images accurately compared to the changing tumor size. Based on RNN and CNN, this Network could present a reliable decision-making method against the traditional clinical methods (48).

Dercle et al. investigated the metastatic CRC response to anti-EGFR treatment in a study. In this study, CT-based radiomic signatures showed a strong relationship between patients' overall survival (OS) and mutational biomarkers. Boundary infiltration and tumor spatial heterogeneity were used as signatures that outperformed other biomarkers, such as KRAS mutations and RECIST 1.1, in predicting sensitivity to anti-EGFR treatments, such as cetuximab. The authors concluded that CT-based radiomic features can effectively predict response to anti-EGFR treatment without requiring invasive and costly operations (49).

A radiomic model was developed to predict the response to HER2-targeted therapy in CRC patients with liver metastasis. This study used CT scan images to divide patients into non-responders if their largest diameter increased by more than 10% after three months of treatment and otherwise responders ($R^+$). The trained algorithm showed a per-patient sensitivity of 92% and specificity of 86% in detecting responding lesions. Therefore, this algorithm could accurately identify the non-responsive patients to whom other possible treatment strategies could be applied (50).

## 5. Radiomics in Lymph node and hepatic metastasis

Metastatic lesions in CRC could affect CRC patients' survival and quality of life. Detecting the real metastatic lesions at the first stages of the disease could enhance the clinician's decision-making. Machine learning and AI could help clinicians find real metastatic lesions with significant accuracy. Several radiomics studies have found that metastatic lesions exhibit high heterogeneity, entropy, and variance, which can be explained by vascularization necrosis and cell clones. These data could be processed with artificial intelligence faster and more accurately than traditional methods (51, 52).

Moreover, texture features extracted from MRI images and clinical features can distinguish metastatic lesions and different types of liver lesions with reasonable specificity and sensitivity rate (51, 53). Consistently, studies have indicated the improved efficacy of CRC management and liver metastasis diagnosis after AI admission. Although machines could not play clinicians' roles in disease management, combining CNN, CT features, clinical data, and AI could increase the efficacy and reliability of metastasis diagnosis and personalized treatment (54).

A study evaluated 120 CRC patients at stage II-III for hepatic metastasis development. Radiomics models examined CT images to detect liver parenchyma. Radiomics analysis results showed heterogeneity in the liver as a protective factor against metastasis (55). Compared to only clinical features, combining radiomics and clinical data is more effective in predicting liver metastasis at the early stages. To predict liver metastasis, CT images of 91 CRC patients were analyzed by Bayesian-optimized random forest with wrapper feature selection. The results showed an AUC of 71% and 86% for clinical features and combination methods, respectively (56).

A study proved the potential of texture analysis of [18F]FDG PET/CT images in predicting metastasis progression of CRC patients. In this study, 52 CRC patients were evaluated for liver metastasis progression after their first line of chemotherapy. This study showed the ability of ML and radiomics to adapt in restaging metastatic CRC with [18F]FDG PET/CT (57). Consistently, a study proved texture analysis's efficacy in diagnosing liver metastatic CRC. MRI imaging was used to evaluate clinical outcomes, recurrence mutational status, etc. Using ML methods could present high efficacy in predicting liver metastasis outcomes, and the best performance was achieved when they adapted KNN for discriminating tumor budding in MRI imaging, in combination with pathological features. In this study, texture analysis could improve the detection of liver metastasis recurrence and recognize the mucinous types of tumors based on clinical and paraclinical findings (58).

In another study, a clinical-radiomics model based on key clinical features and CT could present a great performance in predicting peritoneal metastasis in CRC patients. This study used the LASSO method to extract features with an AUC of 0.78 (59). Similarly, another nomogram based on clinical data and CT-radiomics was developed to predict liver metastasis in CRC patients. The AUC of the nomogram was 0.906 for the cross-validation set and 0.899 for the test set (60). This study indicates the accuracy and efficacy of clinical and radiomics combination nomograms in liver metastasis classification and prediction in CRC patients.

Another study proved the potential of clinical-radiomics nomograms in predicting indeterminate lung lesions. This nomogram used information on CT images and features created by the selection operator algorithm and LASSO. The accuracy and responsibility of the nomogram revealed good radiomics performances for metastasis lesions prediction (61). The texture-radiomics method based on preoperative CT scans has also presented a significant ability for lymph node metastasis prediction. The clinical diagnostic criteria in a study showed a significantly lower accuracy than the texture analyzing method. The clinical diagnostic criteria based on the images' lymph node size could only present 64.8% accuracy, while the texture-radiomics could improve the prediction accuracy up to 81% (62).

Free survival after operations or metastasis ablations is associated with the local metastasis progression. In a study, the radiomics method could predict early post-ablation metastasis progression based on CT images of before and after thermal ablation, and clinical models in combination with radiomics models indicated reliable results for predicting local metastasis progression compared to the radiomics or clinical models individuals (63).

Lymph node metastasis (LNM) is an important factor in the prognosis and treatment of colorectal cancers. The early diagnosis of LNM could significantly affect the management of CRC patients. In a study, Li et al. evaluated the prediction of LNM in colorectal cancer patients by using both clinical manifestations and radiomics analysis features. They concluded that combining clinical data and radiomics features can be reliable for predicting LNM possibility before any invasive operation (64). Another study has shown the accuracy of combining clinical data and radiomics nomograms in predicting lung metastasis in colorectal cancer (61).

Another study used a nomogram based on a logistic regression model to show the efficacy of ML in the pre-treatment prediction of regional lymph node metastasis in CRC patients. The LASSO algorithm was used to extract preoperative 18F-FDG PET/CT diagnostic imaging features. The best performance was achieved by the XGBoost model and the logistic regression (65). In another study, authors utilized CT-based radiomic features to diagnose liver metastasis at the primary undetectable stages. This study revealed a high accuracy (AUC=0.93) in a small clinical sample (66). In another study, 326 CRC-confirmed patients were enrolled in a retrospective study and were evaluated for LN metastasis prediction. This study presented a nomogram based on clinical risk factors, CT reports, radiomics signature, and LN status, which could conveniently predict LN metastasis in CRC patients (67).

In a study, 32 CRC patients' tumor regions were labeled manually and analyzed by Python software to predict the risk of metastasis. The decision tree analyzed Important parameters, and six rules were expressed. Based on analyzing these rules, researchers could classify high-risk and low-risk groups based on the blood volume of the tumor, flow extraction of the nodules, and shape elongation. A combination of texture analysis and functional parameters indicated reasonable results for metastasis classification and T-staging based on preoperative CT scans (68).

## 6. Radiomics in CRC prognosis and survival

Appropriate prediction of prognosis and survival is crucial for proper management and clinical decision-making in CRC. Many studies have used radiomic-based features to predict the survival and outcome of CRC patients after chemotherapy/surgery. Studies have shown the correlation between CT-imaging findings and OS, homogeneity, or heterogeneity in textures of normal liver tissues. In a study by Lubner et al., they defined heterogeneity as skewness, entropy, and mean positive pixels (MPP). Skewness and entropy were negatively correlated with KRAS mutations and survival. It was also demonstrated that high MPP and lower entropy were associated with higher tumor grades in CT images. Moreover, there was a correlation between tumor entropy and OS at coarse filters (69). According to a study that combined bevacizumab and FOLFIRI for chemotherapy of liver metastasis lesions in colorectal cancers, reduction in the sum of lesions kurtosis contributed to the improvement in OS (42).

Moreover, nomograms combining radiomic features with clinical manifestations or pathology images could be used to improve the efficacy and accuracy of outcome predictions. A study evaluated the efficacy of a combined nomogram including radiomics, immunology scoring based on the density of $CD3^+$ and $CD8^+$ cells in tumors, and pathology feature sections to predict OS and postoperative outcomes of colorectal patients with lung metastasis. The result was favorable in predicting disease-free survival (DFS) (AUC=0.875) and OS (AUC=0.860) (70).

In a retrospective study, 1-year survival prediction of CRLM patients was investigated based on pre-treatment CT images. mCRC patient outcome was significantly related to the whole liver tumor burden (WLTB) scores. This study shows that the geometric metastatic spread (GMS) model and radiomics prior model based on WLTB have a significantly higher ability to evaluate the prognosis of mCRC patients. They achieved a significantly higher AUC using machine learning than the clinical models alone (AUC: 0.73 for GMS of WLTB and 0.76 for ARP of WLTB ) (71).

A study was conducted to predict the recurrence of CRC in individuals with stage II/III following surgery. The disease-free survival was evaluated using a contrast-enhanced CT scan. Researchers adapted three ML models to this study: SVM, LASSO, and random forest (RF). The outcomes demonstrated that employing radiomics models might enhance prognostic assessments compared to traditional patient-following methods (p=0.00021) (72). In a retrospective study, 161 CRC patients in the stage I-III of the disease were evaluated for tumor heterogeneity based on CT images. Using the radiomics model showed significantly reliable results for predicting the overall survival of CRC patients. The C-index for the validating and training cohort were 0.678 and 0.780, respectively (73). Another study included 571 CRC patients with stage II-III evaluated for disease-free survival (DFS) with multidetector computed tomography (MDCT) modality. This study used radiomics and DL methods in combination with clinical staging and pathological information, which could reasonably predict the DFS of CRC patients (74).

A study was performed to predict the relapse-free OS of stage I-III CRC patients. In this study, 701 patients were evaluated using their CT images by LASSO Cox to achieve mortality and relapse-free survival. The results showed that radiomics signatures could significantly improve clinical decision-making (AUC for relapse-free survival: 0.744 and 0.768 for overall survival of patients) (75). In another study, researchers used CT images of patients assessed by the LASSO model to achieve the Rad-score for predicting CRC patient prognosis. A combination of nomograms using the Rad-score and TNM stage. Results demonstrated that the Rad-score may enhance CRC patient staging accuracy. The radiomics-based approach evaluated preoperative patients as a non-invasive biomarker, which aids therapeutic intervention in making a decision based on clinicopathological and preoperative CT features nomogram (76).

A study showed that tumor staging information combined with C-X-C Motif Chemokine Ligand 8 (CXCL8)-derived radiomics could enhance prognosis prediction in CRC patients. This study included 141 CRC patients evaluated by preoperative CT imaging. The result of this study demonstrated the appropriate performance of CXCL8-derived radiomics in clinicians' decisions in predicting CRC patients' prognosis (77).

Consistently, 99 mCRC patients treated with palliative first and third-line treatment were evaluated with [18F]FDG PET/ (CT) for their tumor features. The study's results demonstrated the correlation between the decrease in volume of pre-treatment tumors, clinical outcomes, sphericity, and tumor heterogeneity, which were found in [18F]FDG PET image analysis. The authors concluded that good radiomic performance could enhance future disease management and improve PET scan efficiency (78). In another study, pre-treatment [18F]FDG PET was analyzed by a random survival (RSF) model in combination with clinical and pathological features of imaging, which could effectively help in the prediction of CRC patient prognosis (79).

According to studies, the presence of genetic mutations in colorectal cancer can significantly affect the outcome of the patient, including the survival rate, as well as the response of cancer to surgery and chemotherapy (80). Moreover, MSI has been seen in 15 percent of gene instabilities of colorectal cancer patients (81). MSI could show particular biologic features, including a better response to immunotherapy, resistance of the tumor to fluorouracil, and better prognosis (82). The only way to detect gene mutations in colorectal cancers is through invasive and expensive procedures, such as surgery and tissue biopsy. In the past few years, researchers have discovered that AI can improve our understanding of gene mutations by analyzing tumor morphological changes based on their mutations. It is estimated that 35-45% of CRC cases carry mutations of the Kirsten rat sarcoma viral oncogene homolog (KRAS) gene. The type of mutation can influence treatment decisions in patients. The overall survival time has been shortened among patients with specific types of mutations, including BRAF and RAS (83).

BRAF mutation in colorectal cancer patients presents a poor prognosis, lack of efficacy of chemotherapy agents, and poor response to anti-EGFR treatment. In a study, the authors investigated the correlation of radiomics texture analysis of severe CRC patients with 5-year overall survival. This study showed that radiomics texture analysis of CT images of advanced stages of tumors could predict BRAF mutant tumors and 5-year overall survival of CRC patients (84).

Colorectal cancer patients with KRAS gene mutation could affect our therapeutic management. In a retrospective study based on CT scan images, the authors accurately detected KRAS mutation based on CT images and the ResNet model, with an AUC of 0.90. The authors concluded that DL models developed based on pre-treatment CT imaging have a reasonable potential for finding CRC patients with oncogene mutations, and this could reduce the need for invasive procedures, including biopsies (85, 86).

Contrast Enhanced-Magnetic Resonance Imaging (CE-MRI) is another imaging technique that is widely used for detection of liver metastasis in colorectal cancers. In a study using CE-MRI in liver metastases data for detecting RAS mutation status, the authors found that in combination with multivariate analysis, the CE-MRI texture parameters were able to stratify patients with a specificity of 83.3% and the sensitivity of 91.7% *(Figure 4)* (87). Moreover, genetic analysis could improve prognostic information for CRC patients. In a study, the combination of pathological features with gene analysis radiomics, also called radiogenomics, increased the accuracy of prognosis prediction. In this study, ABCC2 expression and clinical stages could predict progression-free survival, while the volume/surface of the tumor and Aldehyde dehydrogenase one family, member A1 (ALDH1A1) expression could predict OS. Therefore, combining radiogenomics and clinical staging information could improve pre-treatment decision-making in the future (88).

Based on MSI data, preoperative CT estimates the CRC patient's survival. In a study, the radiomics technique obtained substantial results in internal and external validation cohorts, with AUCs of 0.77 and 0.78, respectively (89). Radiomics nomograms based on CRC patients' CT images could reliably and accurately predict KRAS mutation. A study used LASSO to develop a nomogram for the survival of patients with MSI. The AUC of this nomogram was 0.92, which was more effective than the 3D/2D radiomics methods (90).

## 7. Challenges and limits

Despite the progression in tumor classification, staging, predicting response to treatment, and predating outcomes of patients with CRC, radiomic-based studies still confront several challenges between institutions. The first is the need for standard protocols for imaging and radiomics workflow. To solve this problem, standardized initiatives and protocols, such as the image biomarker standardization initiative, have been proposed (91). The other challenge is the low quality of radiomics studies. Radiomics quality score (RQS) and quality assessment of diagnostic accuracy studies (QUADAS) are the two most used methods for assessing radiomics study quality. A systematic review by Sanduleanu et al. reported that most included studies had a quality of less than 50% (92). Consistently, in another systematic review of radiomics studies, only 13/76 studies had appropriate quality (5). Heterogeneity in the studied factors and reporting the results is another challenge in radiomics studies (5). The development of standardized methods for including the factors and reporting the results could be an appropriate solution for this problem and ease the understanding and comparison of studies. Type I error and overfitting are other challenges in radiomics studies. To solve this problem, feature reduction can be applied by deleting the number of unnecessary features (93).

On the other hand, almost all radiomics studies face the reproducibility challenge. The reproducibility of radiomics can be challenging, specifically in small datasets related to the nature of images, including image reconstruction parameters, image-acquisition technique, the models used, and feature selection by the user. Furthermore, the lack of external validation is another obstacle in most radiomics studies. To verify the generalizability and reproducibility of radiomics models, they must validate an external dataset (5, 93, 94).

Another challenge is the high dimensionality of the radiomics, which is defined by a large number of measurements compared to a small number of patients. In other words, the large number of features, also named large predictors, and the low number of patients, named "small n-to-p," cause high dimensionality. To obtain reliable results, studies should include a small number of features for a larger number of patients, reducing the overfitting problem (95). Most radiomics studies on CRC have used single-sequence imaging modalities. Since different sequences provide more detailed data about tumors, muti-modality-based radiomics may result in more accurate results. However, this could be time-consuming and is not recommended by some studies (5). In recent years, DL-based models have been frequently used to build radiomics models. DL networks, such as CNN, U.net, ensemble, etc., have been used (96, 97). The growing field of DL and ML could provide appropriate development of more accurte models in the future.

The application of radiomics alone for building models can yield results with low accuracy. To improve accuracy, radiomics can be combined with patients' clinical information. Several studies have reported the higher performance of nomograms combining radiomics data with clinical information in staging (98), detection of MSI (99), preoperative prediction of LN metastasis (100, 101), and prediction of outcome (102) of colorectal cancer. Moreover, other clinical data, such as pathomics, genomics, and immunoscores can also be incorporated to achieve better accuracy (96).

The segmentation of colorectal cancer is a clinical challenge due to heterogeneity in its morphology and boundaries, per-patient diversity in the size and shape of colorectal tissues, and artifacts caused by bowel movements. Moreover, manual segmentation methods are tedious and time-consuming and should be conducted by an experienced expert radiologist. Therefore, automatic segmentation methods have been considered recently. Automatic segmentation models could be more accurate, less time-consuming, and applicable to large datasets (103, 104). Using radiomics tumor imaging might cut expenses and individual errors compared to manual approaches. Using ResNet50 and ResNet152-V2 deep learning models may increase the detection accuracy of malignant ulcers (105).

# Conclusion

In recent years, radiomics-based AI models have opened a primisng window in CRC diagnosis and management. MRI-based radiomics greatly predict survival and treatment outome in rectal cancer (5). In some cases, AI models have consistently outperformed radiologists in the per-patient prediction of lymph node metastasis (106). However, to achieve the reliable clinical application of radiomics-based AI models, some challenges, such as heterogeneity of radiomics studies, overfitting, and dimensionality, still need to be resolved. The future development of AI models in accordance with radiomcis can potentially improve the diagnosis and patient care of colorectal cancer patients.

# Disclosure of conflict of interests

The authors declare that they have no financial or scientific conflicts of interest.

# Funding statement

The authors have not received any funding for this article.

# References

1.	Bray F, Ferlay J, Soerjomataram I, Siegel RL, Torre LA, Jemal A. Global cancer statistics 2018: GLOBOCAN estimates of incidence and mortality worldwide for 36 cancers in 185 countries. CA Cancer J Clin. 2018;68(6):394-424.
2.	Ferlay J, Colombet M, Soerjomataram I, Dyba T, Randi G, Bettio M, et al. Cancer incidence and mortality patterns in Europe: Estimates for 40 countries and 25 major cancers in 2018. Eur J Cancer. 2018;103:356-87.
3.	Biller LH, Schrag D. Diagnosis and Treatment of Metastatic Colorectal Cancer: A Review. Jama. 2021;325(7):669-85.
4.	Anari PY, Lay N, Gopal N, Chaurasia A, Samimi S, Harmon S, et al. An MRI-based radiomics model to predict clear cell renal cell carcinoma growth rate classes in patients with von Hippel-Lindau syndrome. Abdom Radiol (NY). 2022;47(10):3554-62.
5.	Staal FCR, van der Reijd DJ, Taghavi M, Lambregts DMJ, Beets-Tan RGH, Maas M. Radiomics for the Prediction of Treatment Outcome and Survival in Patients With Colorectal Cancer: A Systematic Review. Clin Colorectal Cancer. 2021;20(1):52-71.
6.	Wang Y, Ma LY, Yin XP, Gao BL. Radiomics and Radiogenomics in Evaluation of Colorectal Cancer Liver Metastasis. Front Oncol. 2021;11:689509.
7.	Staal FC, van der Reijd DJ, Taghavi M, Lambregts DM, Beets-Tan RG, Maas M. Radiomics for the prediction of treatment outcome and survival in patients with colorectal cancer: a systematic review. Clinical colorectal cancer. 2021;20(1):52-71.
8.	Agnes SA, Anitha J, Pandian SIA, Peter JD. Classification of Mammogram Images Using Multiscale all Convolutional Neural Network (MA-CNN). J Med Syst. 2019;44(1):30.
9.	Becker AS, Schneider MA, Wurnig MC, Wagner M, Clavien PA, Boss A. Radiomics of liver MRI predict metastases in mice. European radiology experimental. 2018;2(1):1-10.
10.	Huang X, Cheng Z, Huang Y, Liang C, He L, Ma Z, et al. CT-based radiomics signature to discriminate high-grade from low-grade colorectal adenocarcinoma. Academic radiology. 2018;25(10):1285-97.

11. Caruso D, Polici M, Zerunian M, Del Gaudio A, Parri E, Giallorenzi MA, et al. Radiomic Cancer Hallmarks to Identify High-Risk Patients in Non-Metastatic Colon Cancer. Cancers. 2022;14(14):3438.
12. Mascarenhas M, Ribeiro T, Afonso J, Ferreira JP, Cardoso H, Andrade P, et al. Deep learning and colon capsule endoscopy: automatic detection of blood and colonic mucosal lesions using a convolutional neural network. Endoscopy International Open. 2022;10(02):E171-E7.
13. Granata V, Fusco R, De Muzio F, Cutolo C, Setola SV, Dell'Aversana F, et al. Radiomics and machine learning analysis based on magnetic resonance imaging in the assessment of liver mucinous colorectal metastases. La radiologia medica. 2022:1-10.
14. Li Y, Eresen A, Shangguan J, Yang J, Benson AB, 3rd, Yaghmai V, et al. Preoperative prediction of perineural invasion and KRAS mutation in colon cancer using machine learning. J Cancer Res Clin Oncol. 2020;146(12):3165-74.
15. Huang Y, He L, Dong D, Yang C, Liang C, Chen X, et al. Individualized prediction of perineural invasion in colorectal cancer: development and validation of a radiomics prediction model. Chin J Cancer Res. 2018;30(1):40-50.
16. Teo NZ, Wijaya R, Ngu JC. Management of malignant colonic polyps. J Gastrointest Oncol. 2020;11(3):469-74.
17. Grosu S, Wesp P, Graser A, Maurus S, Schulz C, Knösel T, et al. Machine Learning–based differentiation of benign and premalignant colorectal polyps detected with CT colonography in an asymptomatic screening population: A proof-of-concept study. Radiology. 2021;299(2):326-35.
18. Liew WS, Tang TB, Lin C-H, Lu C-K. Automatic colonic polyp detection using integration of modified deep residual convolutional neural network and ensemble learning approaches. Computer Methods and Programs in Biomedicine. 2021;206:106114.
19. Näppi JJ, Pickhardt P, Kim DH, Hironaka T, Yoshida H, editors. Deep learning of contrast-coated serrated polyps for computer-aided detection in CT colonography. Medical Imaging 2017: Computer-Aided Diagnosis; 2017: SPIE.
20. Liang C, Huang Y, He L, Chen X, Ma Z, Dong D, et al. The development and validation of a CT-based radiomics signature for the preoperative discrimination of stage I-II and stage III-IV colorectal cancer. Oncotarget. 2016;7(21):31401.
21. González-Castro V, Cernadas E, Huelga E, Fernández-Delgado M, Porto J, Antunez JR, et al. CT radiomics in colorectal cancer: Detection of KRAS mutation using texture analysis and machine learning. Applied Sciences. 2020;10(18):6214.
22. Wu J, Zhang Q, Zhao Y, Liu Y, Chen A, Li X, et al. Radiomics analysis of iodine-based material decomposition images with dual-energy computed tomography imaging for preoperatively predicting microsatellite instability status in colorectal cancer. Frontiers in oncology. 2019;9:1250.
23. Li Z, Zhong Q, Zhang L, Wang M, Xiao W, Cui F, et al. Computed Tomography-Based Radiomics Model to Preoperatively Predict Microsatellite Instability Status in Colorectal Cancer: A Multicenter Study. Frontiers in Oncology. 2021;11:2563.
24. Fan S, Li X, Cui X, Zheng L, Ren X, Ma W, et al. Computed tomography-based radiomic features could potentially predict microsatellite instability status in stage II colorectal cancer: a preliminary study. Academic radiology. 2019;26(12):1633-40.
25. Golia Pernicka JS, Gagniere J, Chakraborty J, Yamashita R, Nardo L, Creasy JM, et al. Radiomics-based prediction of microsatellite instability in colorectal cancer at initial computed tomography evaluation. Abdominal Radiology. 2019;44(11):3755-63.
26. Pei Q, Yi X, Chen C, Pang P, Fu Y, Lei G, et al. Pre-treatment CT-based radiomics nomogram for predicting microsatellite instability status in colorectal cancer. European Radiology. 2022;32(1):714-24.
27. Hong EK, Bodalal Z, Landolfi F, Bogveradze N, Bos P, Park SJ, et al. Identifying high-risk colon cancer on CT an a radiomics signature improve radiologist's performance for T staging? Abdominal Radiology. 2022:1-8.
28. Bedrikovetski S, Dudi-Venkata NN, Kroon HM, Seow W, Vather R, Carneiro G, et al. Artificial intelligence for pre-operative lymph node staging in colorectal cancer: a systematic review and meta-analysis. BMC cancer. 2021;21:1-10.
29. Granata V, Fusco R, Barretta ML, Picone C, Avallone A, Belli A, et al. Radiomics in hepatic metastasis by colorectal cancer. Infectious Agents and Cancer. 2021;16(1):1-9.

30. Wei J, Cheng J, Gu D, Chai F, Hong N, Wang Y, et al. Deep learning-based radiomics predicts response to chemotherapy in colorectal liver metastases. Medical Physics. 2021;48(1):513-22.
31. Bourbonne V, Schick U, Pradier O, Visvikis D, Metges J-P, Badic B. Radiomics Approaches for the Prediction of Pathological Complete Response after Neoadjuvant Treatment in Locally Advanced Rectal Cancer: Ready for Prime Time? Cancers [Internet]. 2023; 15(2).
32. Chen BY, Xie H, Li Y, Jiang XH, Xiong L, Tang XF, et al. MRI-Based Radiomics Features to Predict Treatment Response to Neoadjuvant Chemotherapy in Locally Advanced Rectal Cancer: A Single Center, Prospective Study. Front Oncol. 2022;12:801743.
33. Popat S, Hubner R, Houlston R. Systematic review of microsatellite instability and colorectal cancer prognosis. Journal of clinical oncology. 2005;23(3):609-18.
34. Liu X, Ran R, Shao B, Rugo HS, Yang Y, Hu Z, et al. Combined peripheral natural killer cell and circulating tumor cell enumeration enhance prognostic efficiency in patients with metastatic triple-negative breast cancer. Chinese Journal of Cancer Research. 2018;30(3):315.
35. Li Z, Zhang J, Zhong Q, Feng Z, Shi Y, Xu L, et al. Development and external validation of a multiparametric MRI-based radiomics model for preoperative prediction of microsatellite instability status in rectal cancer: a retrospective multicenter study. European Radiology. 2022:1-9.
36. Giannini V, Pusceddu L, Defeudis A, Nicoletti G, Cappello G, Mazzetti S, et al. Delta-Radiomics Predicts Response to First-Line Oxaliplatin-Based Chemotherapy in Colorectal Cancer Patients with Liver Metastases. Cancers (Basel). 2022;14(1).
37. Yan H, Yu TN. Radiomics-clinical nomogram for response to chemotherapy in synchronous liver metastasis of colorectal cancer: Good, but not good enough. World J Gastroenterol. 2022;28(9):973-5.
38. Ahn SJ, Kim JH, Park SJ, Han JK. Prediction of the therapeutic response after FOLFOX and FOLFIRI treatment for patients with liver metastasis from colorectal cancer using computerized CT texture analysis. European journal of radiology. 2016;85(10):1867-74.
39. Andersen IR, Thorup K, Andersen MB, Olesen R, Mortensen FV, Nielsen DT, et al. Texture in the monitoring of regorafenib therapy in patients with colorectal liver metastases. Acta Radiologica. 2019;60(9):1084-93.
40. Zhang H, Li W, Hu F, Sun Y, Hu T, Tong T. MR texture analysis: potential imaging biomarker for predicting the chemotherapeutic response of patients with colorectal liver metastases. Abdominal Radiology. 2019;44(1):65-71.
41. Wei J, Cheng J, Gu D, Chai F, Hong N, Wang Y, et al. Deep learning-based radiomics predicts response to chemotherapy in colorectal liver metastases. Med Phys. 2021;48(1):513-22.
42. Dohan A, Gallix B, Guiu B, Le Malicot K, Reinhold C, Soyer P, et al. Early evaluation using a radiomic signature of unresectable hepatic metastases to predict outcome in patients with colorectal cancer treated with FOLFIRI and bevacizumab. Gut. 2020;69(3):531-9.
43. Ma Y-Q, Wen Y, Liang H, Zhong J-G, Pang P-P. Magnetic resonance imaging-radiomics evaluation of response to chemotherapy for synchronous liver metastasis of colorectal cancer. World Journal of Gastroenterology. 2021;27(38):6465.
44. Alvarez-Jimenez C, Antunes JT, Talasila N, Bera K, Brady JT, Gollamudi J, et al. Radiomic texture and shape descriptors of the rectal environment on post-chemoradiation T2-weighted MRI are associated with pathologic tumor stage regression in rectal cancers: a retrospective, multi-institution study. Cancers. 2020;12(8):2027.
45. Hasegawa T, Takaki H, Kodama H, Yamanaka T, Nakatsuka A, Sato Y, et al. Three-year Survival Rate after Radiofrequency Ablation for Surgically Resectable Colorectal Lung Metastases: A Prospective Multicenter Study. Radiology. 2020;294(3):686-95.
46. Markich R, Palussière J, Catena V, Cazayus M, Fonck M, Bechade D, et al. Radiomics complements clinical, radiological, and technical features to assess local control of colorectal cancer lung metastases treated with radiofrequency ablation. European Radiology. 2021;31(11):8302-14.

47. Wu XZ. A new classification system of anticancer drugs - based on cell biological mechanisms. Med Hypotheses. 2006;66(5):883-7.
48. Lu L, Dercle L, Zhao B, Schwartz LH. Deep learning for the prediction of early on-treatment response in metastatic colorectal cancer from serial medical imaging. Nature communications. 2021;12(1):1-11.
49. Dercle L, Lu L, Schwartz LH, Qian M, Tejpar S, Eggleton P, et al. Radiomics response signature for identification of metastatic colorectal cancer sensitive to therapies targeting EGFR pathway. JNCI: Journal of the National Cancer Institute. 2020;112(9):902-12.
50. Giannini V, Rosati S, Defeudis A, Balestra G, Vassallo L, Cappello G, et al. Radiomics predicts response of individual HER2-amplified colorectal cancer liver metastases in patients treated with HER2-targeted therapy. International Journal of Cancer. 2020;147(11):3215-23.
51. Fiz F, Viganò L, Gennaro N, Costa G, La Bella L, Boichuk A, et al. Radiomics of liver metastases: a systematic review. Cancers. 2020;12(10):2881.
52. de la Pinta C, Castillo ME, Collado M, Galindo-Pumariño C, Peña C. Radiogenomics: hunting down liver metastasis in colorectal cancer patients. Cancers. 2021;13(21):5547.
53. Gatos I, Tsantis S, Karamesini M, Spiliopoulos S, Karnabatidis D, Hazle JD, et al. Focal liver lesions segmentation and classification in nonenhanced T2-weighted MRI. Medical physics. 2017;44(7):3695-705.
54. Rompianesi G, Pegoraro F, Ceresa CD, Montalti R, Troisi RI. Artificial intelligence in the diagnosis and management of colorectal cancer liver metastases. World Journal of Gastroenterology. 2022;28(1):108.
55. Creasy JM, Cunanan KM, Chakraborty J, McAuliffe JC, Chou J, Gonen M, et al. Differences in liver parenchyma are measurable with CT radiomics at initial colon resection in patients that develop hepatic metastases from stage II/III colon cancer. Annals of surgical oncology. 2021;28(4):1982-9.
56. Taghavi M, Trebeschi S, Simões R, Meek DB, Beckers RC, Lambregts DM, et al. Machine learning-based analysis of CT radiomics model for prediction of colorectal metachronous liver metastases. Abdominal Radiology. 2021;46(1):249-56.
57. Alongi P, Stefano A, Comelli A, Spataro A, Formica G, Laudicella R, et al. Artificial Intelligence Applications on Restaging [18F] FDG PET/CT in Metastatic Colorectal Cancer: A Preliminary Report of Morpho-Functional Radiomics Classification for Prediction of Disease Outcome. Applied Sciences. 2022;12(6):2941.
58. Granata V, Fusco R, De Muzio F, Cutolo C, Setola SV, dell'Aversana F, et al. Contrast MR-based radiomics and machine learning analysis to assess clinical outcomes following liver resection in colorectal liver metastases: a preliminary study. Cancers. 2022;14(5):1110.
59. Li M, Sun K, Dai W, Xiang W, Zhang Z, Zhang R, et al. Preoperative prediction of peritoneal metastasis in colorectal cancer using a clinical-radiomics model. European journal of radiology. 2020;132:109326.
60. Li M, Li X, Guo Y, Miao Z, Liu X, Guo S, et al. Development and assessment of an individualized nomogram to predict colorectal cancer liver metastases. Quantitative imaging in medicine and surgery. 2020;10(2):397.
61. Hu T, Wang S, Huang L, Wang J, Shi D, Li Y, et al. A clinical-radiomics nomogram for the preoperative prediction of lung metastasis in colorectal cancer patients with indeterminate pulmonary nodules. European radiology. 2019;29(1):439-49.
62. Eresen A, Li Y, Yang J, Shangguan J, Velichko Y, Yaghmai V, et al. Preoperative assessment of lymph node metastasis in Colon Cancer patients using machine learning: a pilot study. Cancer imaging. 2020;20(1):1-9.
63. Taghavi M, Staal F, Gomez Munoz F, Imani F, Meek DB, Simões R, et al. CT-based radiomics analysis before thermal ablation to predict local tumor progression for colorectal liver metastases. Cardiovascular and Interventional Radiology. 2021;44(6):913-20.
64. Li M, Zhang J, Dan Y, Yao Y, Dai W, Cai G, et al. A clinical-radiomics nomogram for the preoperative prediction of lymph node metastasis in colorectal cancer. Journal of translational medicine. 2020;18(1):1-10.

65. He J, Wang Q, Zhang Y, Wu H, Zhou Y, Zhao S. Preoperative prediction of regional lymph node metastasis of colorectal cancer based on 18F-FDG PET/CT and machine learning. Annals of Nuclear Medicine. 2021;35(5):617-27.
66. Rocca A, Brunese MC, Santone A, Avella P, Bianco P, Scacchi A, et al. Early diagnosis of liver metastases from colorectal cancer through CT radiomics and formal methods: a pilot study. Journal of Clinical Medicine. 2022;11(1):31.
67. Huang Y-q, Liang C-h, He L, Tian J, Liang C-s, Chen X, et al. Development and validation of a radiomics nomogram for preoperative prediction of lymph node metastasis in colorectal cancer. Journal of clinical oncology. 2016;34(18):2157-64.
68. Dou Y, Liu Y, Kong X, Yang S. T staging with functional and radiomics parameters of computed tomography in colorectal cancer patients. Medicine. 2022;101(21):e29244.
69. Lubner MG, Stabo N, Lubner SJ, Del Rio AM, Song C, Halberg RB, et al. CT textural analysis of hepatic metastatic colorectal cancer: pre-treatment tumor heterogeneity correlates with pathology and clinical outcomes. Abdominal imaging. 2015;40(7):2331-7.
70. Wang R, Dai W, Gong J, Huang M, Hu T, Li H, et al. Development of a novel combined nomogram model integrating deep learning-pathomics, radiomics and immunoscore to predict postoperative outcome of colorectal cancer lung metastasis patients. Journal of hematology & oncology. 2022;15(1):1-6.
71. Mühlberg A, Holch JW, Heinemann V, Huber T, Moltz J, Maurus S, et al. The relevance of CT-based geometric and radiomics analysis of whole liver tumor burden to predict survival of patients with metastatic colorectal cancer. European Radiology. 2021;31(2):834-46.
72. Badic B, Da-Ano R, Poirot K, Jaouen V, Magnin B, Gagnière J, et al. Prediction of recurrence after surgery in colorectal cancer patients using radiomics from diagnostic contrast-enhanced computed tomography: a two-center study. European Radiology. 2022;32(1):405-14.
73. Huang Y, He L, Li Z, Chen X, Han C, Zhao K, et al. Coupling radiomics analysis of CT image with diversification of tumor ecosystem: A new insight to overall survival in stage I– III colorectal cancer. Chinese Journal of Cancer Research. 2022;34(1):40.
74. Yao X, Sun C, Xiong F, Zhang X, Cheng J, Wang C, et al. Radiomic signature-based nomogram to predict disease-free survival in stage II and III colon cancer. European Journal of Radiology. 2020;131:109205.
75. Dai W, Mo S, Han L, Xiang W, Li M, Wang R, et al. Prognostic and predictive value of radiomics signatures in stage I-III colon cancer. Clinical and translational medicine. 2020;10(1):288-93.
76. Cai D, Duan X, Wang W, Huang Z-P, Zhu Q, Zhong M-E, et al. A metabolism-related radiomics signature for predicting the prognosis of colorectal cancer. Frontiers in molecular biosciences. 2021;7:613918.
77. Chu Y, Li J, Zeng Z, Huang B, Zhao J, Liu Q, et al. A novel model based on cxcl8-derived radiomics for prognosis prediction in colorectal cancer. Frontiers in oncology. 2020;10:575422.
78. Van Helden E, Vacher Y, Van Wieringen W, Van Velden F, Verheul H, Hoekstra O, et al. Radiomics analysis of pre-treatment [18F] FDG PET/CT for patients with metastatic colorectal cancer undergoing palliative systemic treatment. European journal of nuclear medicine and molecular imaging. 2018;45(13):2307-17.
79. Lv L, Xin B, Hao Y, Yang Z, Xu J, Wang L, et al. Radiomic analysis for predicting prognosis of colorectal cancer from preoperative 18F-FDG PET/CT. Journal of translational medicine. 2022;20(1):1-11.
80. Armaghany T, Wilson JD, Chu Q, Mills G. Genetic alterations in colorectal cancer. Gastrointestinal cancer research: GCR. 2012;5(1):19.
81. Søreide K, Janssen E, Söiland H, Körner H, Baak J. Microsatellite instability in colorectal cancer. Journal of British Surgery. 2006;93(4):395-406.
82. Hildebrand LA, Pierce CJ, Dennis M, Paracha M, Maoz A. Artificial intelligence for histology-based detection of microsatellite instability and prediction of response to immunotherapy in colorectal cancer. Cancers. 2021;13(3):391.
83. Shi R, Chen W, Yang B, Qu J, Cheng Y, Zhu Z, et al. Prediction of KRAS, NRAS and BRAF status in colorectal cancer patients with liver metastasis using a deep artificial neural network based on radiomics and semantic features. American journal of cancer research. 2020;10(12):4513.

84. Negreros-Osuna AA, Parakh A, Corcoran RB, Pourvaziri A, Kambadakone A, Ryan DP, et al. Radiomics texture features in advanced colorectal cancer: correlation with BRAF mutation and 5-year overall survival. Radiology: Imaging Cancer. 2020;2(5).
85. He K, Liu X, Li M, Li X, Yang H, Zhang H. Noninvasive KRAS mutation estimation in colorectal cancer using a deep learning method based on CT imaging. BMC medical imaging. 2020;20(1):1-9.
86. Shi R, Chen W, Yang B, Qu J, Cheng Y, Zhu Z, et al. Prediction of KRAS, NRAS and BRAF status in colorectal cancer patients with liver metastasis using a deep artificial neural network based on radiomics and semantic features. Am J Cancer Res. 2020;10(12):4513-26.
87. Granata V, Fusco R, Avallone A, De Stefano A, Ottaiano A, Sbordone C, et al. Radiomics-Derived Data by Contrast Enhanced Magnetic Resonance in RAS Mutations Detection in Colorectal Liver Metastases. Cancers (Basel). 2021;13(3).
88. Badic B, Hatt M, Durand S, Jossic-Corcos CL, Simon B, Visvikis D, et al. Radiogenomics-based cancer prognosis in colorectal cancer. Scientific reports. 2019;9(1):1-7.
89. Chen X, He L, Li Q, Liu L, Li S, Zhang Y, et al. Non-invasive prediction of microsatellite instability in colorectal cancer by a genetic algorithm–enhanced artificial neural network–based CT radiomics signature. European Radiology. 2023;33(1):11-22.
90. Xue T, Peng H, Chen Q, Li M, Duan S, Feng F. Preoperative prediction of KRAS mutation status in colorectal cancer using a CT-based radiomics nomogram. The British Journal of Radiology. 2022;95(1134):20211014.
91. Zwanenburg A, Vallières M, Abdalah MA, Aerts H, Andrearczyk V, Apte A, et al. The Image Biomarker Standardization Initiative: Standardized Quantitative Radiomics for High-Throughput Image-based Phenotyping. Radiology. 2020;295(2):328-38.
92. Sanduleanu S, Woodruff HC, de Jong EEC, van Timmeren JE, Jochems A, Dubois L, et al. Tracking tumor biology with radiomics: A systematic review utilizing a radiomics quality score. Radiother Oncol. 2018;127(3):349-60.
93. Park JE, Park SY, Kim HJ, Kim HS. Reproducibility and Generalizability in Radiomics Modeling: Possible Strategies in Radiologic and Statistical Perspectives. Korean J Radiol. 2019;20(7):1124-37.
94. Hou M, Sun JH. Emerging applications of radiomics in rectal cancer: State of the art and future perspectives. World J Gastroenterol. 2021;27(25):3802-14.
95. Clarke R, Ressom HW, Wang A, Xuan J, Liu MC, Gehan EA, et al. The properties of high-dimensional data spaces: implications for exploring gene and protein expression data. Nat Rev Cancer. 2008;8(1):37-49.
96. Wang R, Dai W, Gong J, Huang M, Hu T, Li H, et al. Development of a novel combined nomogram model integrating deep learning-pathomics, radiomics and immunoscore to predict postoperative outcome of colorectal cancer lung metastasis patients. Journal of Hematology & Oncology. 2022;15(1):11.
97. Zhao J, Wang H, Zhang Y, Wang R, Liu Q, Li J, et al. Deep learning radiomics model related with genomics phenotypes for lymph node metastasis prediction in colorectal cancer. Radiother Oncol. 2022;167:195-202.
98. Lin X, Zhao S, Jiang H, Jia F, Wang G, He B, et al. A radiomics-based nomogram for preoperative T staging prediction of rectal cancer. Abdominal Radiology. 2021;46(10):4525-35.
99. Ying M, Pan J, Lu G, Zhou S, Fu J, Wang Q, et al. Development and validation of a radiomics-based nomogram for the preoperative prediction of microsatellite instability in colorectal cancer. BMC Cancer. 2022;22(1):524.
100. Li M, Zhang J, Dan Y, Yao Y, Dai W, Cai G, et al. A clinical-radiomics nomogram for the preoperative prediction of lymph node metastasis in colorectal cancer. Journal of Translational Medicine. 2020;18(1):46.
101. Cheng Y, Yu Q, Meng W, Jiang W. Clinico-Radiologic Nomogram Using Multiphase CT to Predict Lymph Node Metastasis in Colon Cancer. Mol Imaging Biol. 2022;24(5):798-806.
102. Li M, Xu G, Chen Q, Xue T, Peng H, Wang Y, et al. Computed Tomography-based Radiomics Nomogram for the Preoperative Prediction of Tumor Deposits and Clinical Outcomes in Colon Cancer: a Multicenter Study. Academic Radiology. 2022.
103. Akilandeswari A, Sungeetha D, Joseph C, Thaiyalnayaki K, Baskaran K, Jothi Ramalingam R, et al. Automatic Detection and Segmentation of Colorectal Cancer with

Deep Residual Convolutional Neural Network. Evidence-Based Complementary and Alternative Medicine. 2022;2022:3415603.
104. Akilandeswari A, Sungeetha D, Joseph C, Thaiyalnayaki K, Baskaran K, Jothi Ramalingam R, et al. Automatic Detection and Segmentation of Colorectal Cancer with Deep Residual Convolutional Neural Network. Evid Based Complement Alternat Med. 2022;2022:3415603.
105. Masmoudi Y, Ramzan M, Khan SA, Habib M. Optimal feature extraction and ulcer classification from WCE image data using deep learning. Soft Computing. 2022:1-14.
106. Bedrikovetski S, Dudi-Venkata NN, Kroon HM, Seow W, Vather R, Carneiro G, et al. Artificial intelligence for pre-operative lymph node staging in colorectal cancer: a systematic review and meta-analysis. BMC Cancer. 2021;21(1):1058.
107. Rafaelsen SR, Dam C, Vagn-Hansen C, Møller J, Rahr HB, Sjöström M, et al. CT and 3 Tesla MRI in the TN Staging of Colon Cancer: A Prospective, Blind Study. Current Oncology. 2022;29(2):1069-79.
108. Alvarez-Jimenez C, Antunes JT, Talasila N, Bera K, Brady JT, Gollamudi J, et al. Radiomic Texture and Shape Descriptors of the Rectal Environment on Post-Chemoradiation T2-Weighted MRI are Associated with Pathologic Tumor Stage Regression in Rectal Cancers: A Retrospective, Multi-Institution Study. Cancers (Basel). 2020;12(8).
109. Hong EK, Bodalal Z, Landolfi F, Bogveradze N, Bos P, Park SJ, et al. Identifying high-risk colon cancer on CT an a radiomics signature improve radiologist's performance for T staging? Abdominal Radiology. 2022;47(8):2739-46.
110. Granata V, Fusco R, De Muzio F, Cutolo C, Setola SV, Dell'Aversana F, et al. Radiomics and machine learning analysis based on magnetic resonance imaging in the assessment of liver mucinous colorectal metastases. La radiologia medica. 2022;127(7):763-72.
111. Li Y, Eresen A, Shangguan J, Yang J, Benson AB, Yaghmai V, et al. Preoperative prediction of perineural invasion and KRAS mutation in colon cancer using machine learning. Journal of Cancer Research and Clinical Oncology. 2020;146:3165-74.
112. Huang Y, He L, Dong D, Yang C, Liang C, Chen X, et al. Individualized prediction of perineural invasion in colorectal cancer: development and validation of a radiomics prediction model. Chinese Journal of Cancer Research. 2018;30(1):40.
113. Grosu S, Wesp P, Graser A, Maurus S, Schulz C, Knoesel T, et al. Machine learning–based differentiation of benign and premalignant colorectal polyps detected with CT Colonography in an asymptomatic screening population: a proof-of-concept study. Radiology. 2021;299(2):326-35.
114. Wesp P, Grosu S, Graser A, Maurus S, Schulz C, Knösel T, et al. Deep learning in CT colonography: differentiating premalignant from benign colorectal polyps. European Radiology. 2022;32(7):4749-59.
115. Yuan Z, Xu T, Cai J, Zhao Y, Cao W, Fichera A, et al. Development and validation of an image-based deep learning algorithm for detection of synchronous peritoneal carcinomatosis in colorectal cancer. Annals of surgery. 2022;275(4):e645-e51.
116. Li Z, Zhong Q, Zhang L, Wang M, Xiao W, Cui F, et al. Computed tomography-based radiomics model to preoperatively predict microsatellite instability status in colorectal cancer: a multicenter study. Frontiers in Oncology. 2021;11:666786.
117. Golia Pernicka JS, Gagniere J, Chakraborty J, Yamashita R, Nardo L, Creasy JM, et al. Radiomics-based prediction of microsatellite instability in colorectal cancer at initial computed tomography evaluation. Abdominal Radiology. 2019;44:3755-63.
118. Pei Q, Yi X, Chen C, Pang P, Fu Y, Lei G, et al. Pre-treatment CT-based radiomics nomogram for predicting microsatellite instability status in colorectal cancer. European Radiology. 2022;32:714-24.
119. Giannini V, Pusceddu L, Defeudis A, Nicoletti G, Cappello G, Mazzetti S, et al. Delta-radiomics predicts response to first-line oxaliplatin-based chemotherapy in colorectal cancer patients with liver metastases. Cancers. 2022;14(1):241.
120. Zhang H, Li W, Hu F, Sun Y, Hu T, Tong T. MR texture analysis: potential imaging biomarker for predicting the chemotherapeutic response of patients with colorectal liver metastases. Abdominal Radiology. 2019;44:65-71.
121. Lu L, Dercle L, Zhao B, Schwartz LH. Deep learning for the prediction of early on-treatment response in metastatic colorectal cancer from serial medical imaging. Nature communications. 2021;12(1):6654.

122. Creasy JM, Cunanan KM, Chakraborty J, McAuliffe JC, Chou J, Gonen M, et al. Differences in liver parenchyma are measurable with CT radiomics at initial colon resection in patients that develop hepatic metastases from stage II/III colon cancer. Annals of surgical oncology. 2021;28:1982-9.
123. Taghavi M, Trebeschi S, Simões R, Meek DB, Beckers RC, Lambregts DM, et al. Machine learning-based analysis of CT radiomics model for prediction of colorectal metachronous liver metastases. Abdominal Radiology. 2021;46:249-56.
124. Li M, Sun K, Dai W, Xiang W, Zhang R, Wang R, et al. Preoperative prediction of peritoneal metastasis in colorectal cancer using a clinical-radiomics model. European journal of radiology. 2020;132:109326.
125. Hu T, Wang S, Huang L, Wang J, Shi D, Li Y, et al. A clinical-radiomics nomogram for the preoperative prediction of lung metastasis in colorectal cancer patients with indeterminate pulmonary nodules. European radiology. 2019;29:439-49.
126. Eresen A, Li Y, Yang J, Shangguan J, Velichko Y, Yaghmai V, et al. Preoperative assessment of lymph node metastasis in Colon Cancer patients using machine learning: A pilot study. Cancer imaging. 2020;20:1-9.
127. Taghavi M, Staal F, Gomez Munoz F, Imani F, Meek DB, Simões R, et al. CT-based radiomics analysis before thermal ablation to predict local tumor progression for colorectal liver metastases. Cardiovascular and Interventional Radiology. 2021;44:913-20.
128. He J, Wang Q, Zhang Y, Wu H, Zhou Y, Zhao S. Preoperative prediction of regional lymph node metastasis of colorectal cancer based on 18 F-FDG PET/CT and machine learning. Annals of Nuclear Medicine. 2021;35:617-27.
129. Liu C, Meng Q, Zeng Q, Chen H, Shen Y, Li B, et al. An Exploratory Study on the Stable Radiomics Features of Metastatic Small Pulmonary Nodules in Colorectal Cancer Patients. Frontiers in Oncology. 2021:2579.
130. Rizzetto F, Calderoni F, De Mattia C, Defeudis A, Giannini V, Mazzetti S, et al. Impact of inter-reader contouring variability on textural radiomics of colorectal liver metastases. European radiology experimental. 2020;4:1-12.
131. Bae H, Lee H, Kim S, Han K, Rhee H, Kim D-k, et al. Radiomics analysis of contrast-enhanced CT for classification of hepatic focal lesions in colorectal cancer patients: Its limitations compared to radiologists. European Radiology. 2021;31(11):8786-96.
132. Lubner MG, Stabo N, Lubner SJ, Del Rio AM, Song C, Halberg RB, et al. CT textural analysis of hepatic metastatic colorectal cancer: pre-treatment tumor heterogeneity correlates with pathology and clinical outcomes. Abdominal imaging. 2015;40:2331-7.
133. Kang J, Lee J-H, Lee HS, Cho E-S, Park EJ, Baik SH, et al. Radiomics features of 18f-fluorodeoxyglucose positron-emission tomography as a novel prognostic signature in colorectal cancer. Cancers. 2021;13(3):392.
134. Wang R, Dai W, Gong J, Huang M, Hu T, Li H, et al. Development of a novel combined nomogram model integrating deep learning-pathomics, radiomics and immunoscore to predict postoperative outcome of colorectal cancer lung metastasis patients. Journal of hematology & oncology. 2022;15(1):11.
135. Mühlberg A, Holch JW, Heinemann V, Huber T, Moltz J, Maurus S, et al. The relevance of CT-based geometric and radiomics analysis of whole liver tumor burden to predict survival of patients with metastatic colorectal cancer. European Radiology. 2021;31:834-46.
136. Van Helden E, Vacher Y, Van Wieringen W, Van Velden F, Verheul H, Hoekstra O, et al. Radiomics analysis of pre-treatment [18 F] FDG PET/CT for patients with metastatic colorectal cancer undergoing palliative systemic treatment. European Journal of Nuclear Medicine and Molecular Imaging. 2018;45:2307-17.
137. Lv L, Xin B, Hao Y, Yang Z, Xu J, Wang L, et al. Radiomic analysis for predicting prognosis of colorectal cancer from preoperative 18F-FDG PET/CT. Journal of Translational Medicine. 2022;20(1):66.
138. He K, Liu X, Li M, Li X, Yang H, Zhang H. Noninvasive KRAS mutation estimation in colorectal cancer using a deep learning method based on CT imaging. BMC medical imaging. 2020;20:1-9.
139. Granata V, Fusco R, Avallone A, De Stefano A, Ottaiano A, Sbordone C, et al. Radiomics-derived data by contrast enhanced magnetic resonance in RAS mutations detection in colorectal liver metastases. Cancers. 2021;13(3):453.

140. Badic B, Hatt M, Durand S, Jossic-Corcos CL, Simon B, Visvikis D, et al. Radiogenomics-based cancer prognosis in colorectal cancer. Scientific Reports. 2019;9(1):9743.
141. Chen X, He L, Li Q, Liu L, Li S, Zhang Y, et al. Non-invasive prediction of microsatellite instability in colorectal cancer by a genetic algorithm–enhanced artificial neural network–based CT radiomics signature. European Radiology. 2022:1-12.
142. Huang Y-C, Tsai Y-S, Li C-I, Chan R-H, Yeh Y-M, Chen P-C, et al. Adjusted CT Image-Based Radiomic Features Combined with Immune Genomic Expression Achieve Accurate Prognostic Classification and Identification of Therapeutic Targets in Stage III Colorectal Cancer. Cancers. 2022;14(8):1895.
143. Fan S, Cui X, Liu C, Li X, Zheng L, Song Q, et al. CT-based radiomics signature: a potential biomarker for predicting postoperative recurrence risk in stage II colorectal cancer. Frontiers in Oncology. 2021;11:644933.
144. Song M, Li S, Wang H, Hu K, Wang F, Teng H, et al. MRI radiomics independent of clinical baseline characteristics and neoadjuvant treatment modalities predicts response to neoadjuvant therapy in rectal cancer. British Journal of Cancer. 2022;127(2):249-57.
145. Wang F, Tan BF, Poh SS, Siow TR, Lim FLWT, Yip CSP, et al. Predicting outcomes for locally advanced rectal cancer treated with neoadjuvant chemoradiation with CT-based radiomics. Scientific Reports. 2022;12(1):6167.
146. Feng L, Liu Z, Li C, Li Z, Lou X, Shao L, et al. Development and validation of a radiopathomics model to predict pathological complete response to neoadjuvant chemoradiotherapy in locally advanced rectal cancer: a multicentre observational study. The Lancet Digital Health. 2022;4(1):e8-e17.
147. Dinapoli N, Barbaro B, Gatta R, Chiloiro G, Casà C, Masciocchi C, et al. Magnetic resonance, vendor-independent, intensity histogram analysis predicting pathologic complete response after radiochemotherapy of rectal cancer. International Journal of Radiation Oncology* Biology* Physics. 2018;102(4):765-74.
148. Defeudis A, Mazzetti S, Panic J, Micilotta M, Vassallo L, Giannetto G, et al. MRI-based radiomics to predict response in locally advanced rectal cancer: comparison of manual and automatic segmentation on external validation in a multicentre study. European Radiology Experimental. 2022;6(1):19.

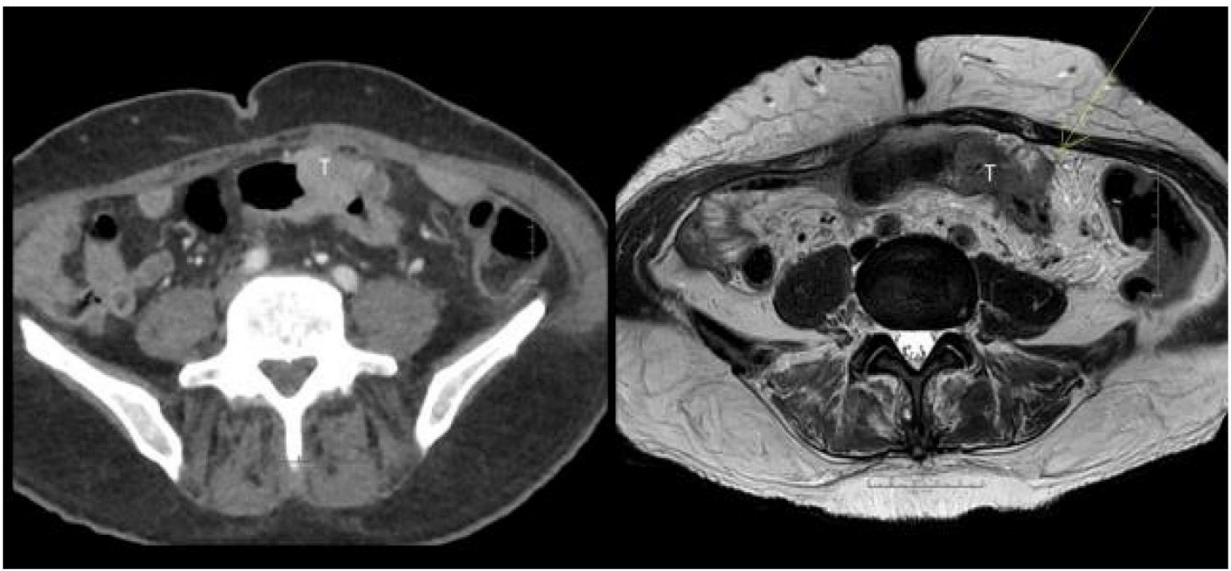

**Figure 1. left:** CT image of a tumor (T) in an elongated sigmoid sling, reported as a large T2,N0,V0 tumor. **Right**: MRI at 3T reported a T3c,N1,V2 tumor with an outgrowth from the colonic wall of 7 mm (arrow). Histopathology confirmed all the MRI findings (107).

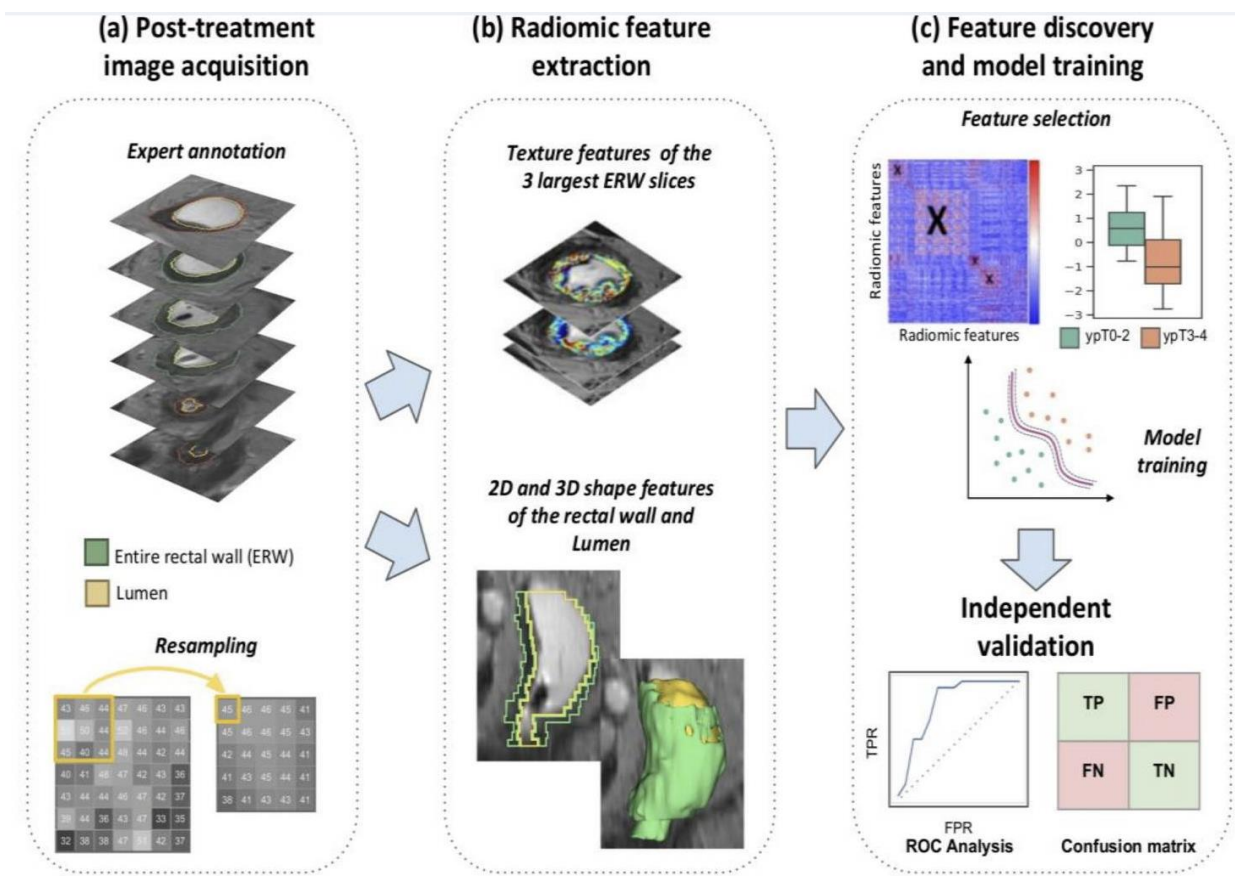

**Figure 2.** Example of a radiomics workflowOverview of radiomics workflow for evaluating pathologic tumor stage regression using post-treatment T2w MRI (108).

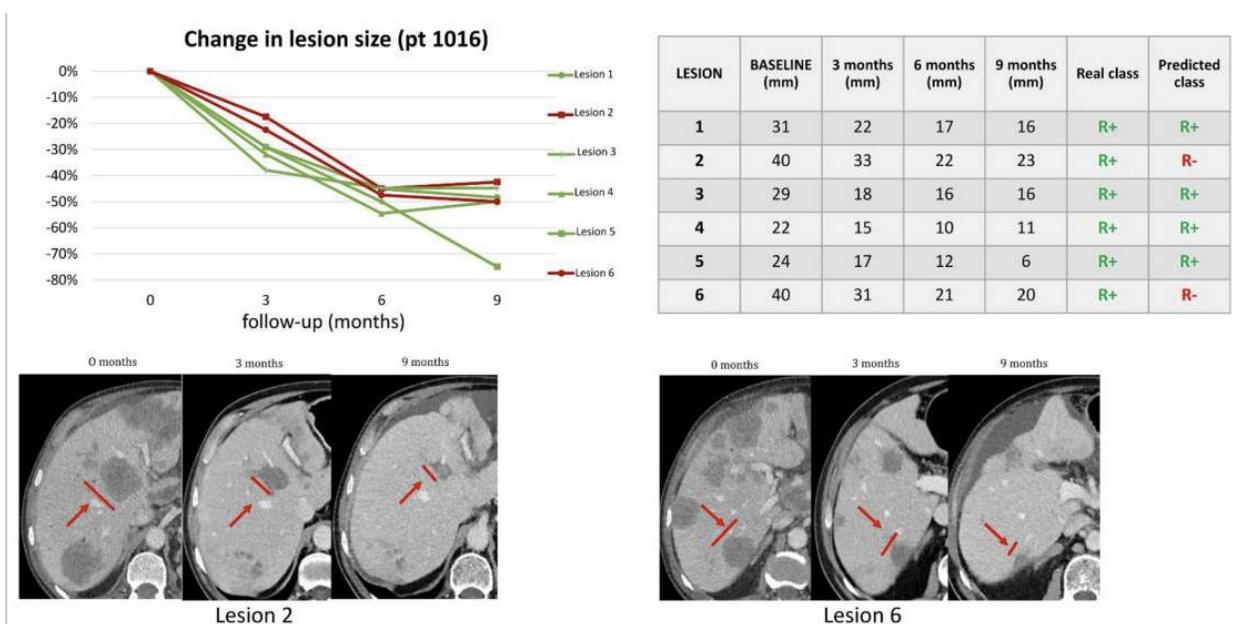

**Figure 3.** Example of prediction of response to first-line Oxaliplatin-based chemotherapy in a colorectal cancer patient with liver metastases using Delta-Radiomics. The algorithm correctly classified a R+ 4/6 lesions while 2/6 lesions were misclassified as R−. The red line in the graph represents the metastasis that went into progression before 8 months, while the green lines represent metastases that responded for at least 8 months. The table lists the patient's liver

| covariate-adjusted text Author | Imaging modality | Rdiomics features | AUC/accuracy | Findings |
|---|---|---|---|---|

metastases, size at baseline, and subsequent timepoints. The real and predicted class columns show respectively the 8-month response of each lesion based on size variations and the classification as predicted by the classifiers (36).

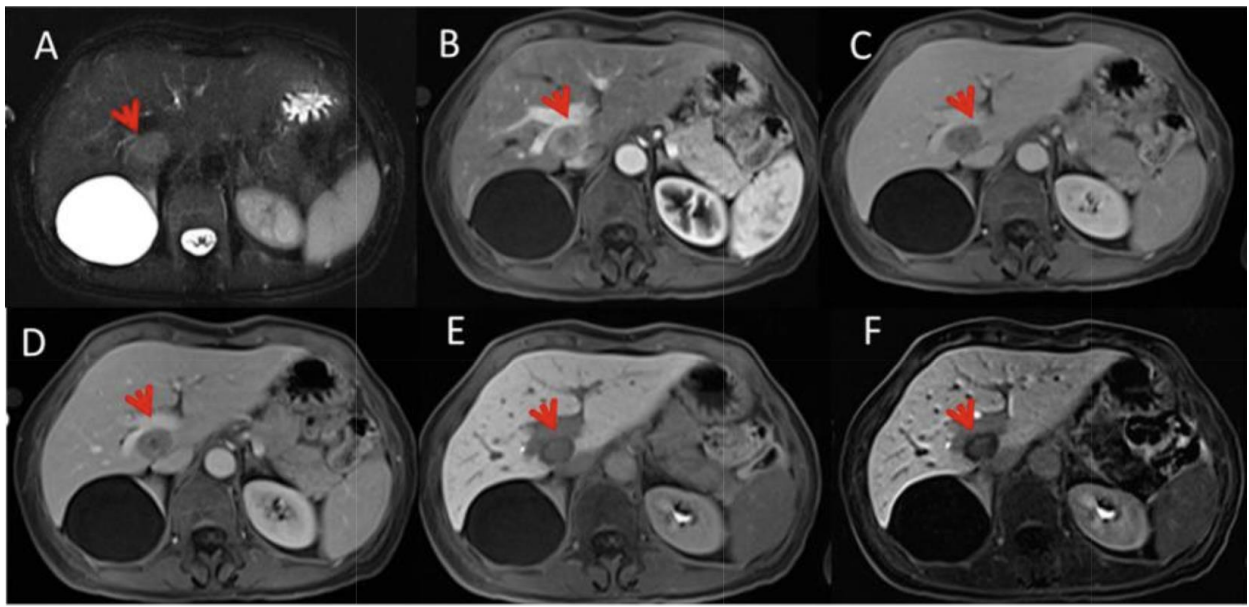

**Figure 4.** Examples of liver metastasis in rectal cancer in a 63 year-old male patient on MRI. In (**A**) (TRUFISP T2-W fat sat) the lesion (arrow) is hyperintense. Contrast study with Gd-EOB-DTPA: the lesion (arrow) shows a peripheral rim enhancement with a hypointense core, in arterial phase (**B**). In the portal (**C**) and transitional phase (**D**), the lesion is hypointense. In the HPB phase (**E**,**F**), the lesion is hypointense, with a target appearance (87).

|  | **Author** | **Imaging modality** | **Rdiomics features** | **AUC/accuracy** | **Findings** |
|---|---|---|---|---|---|
| **CRC diagnosis** | Hong et al. (109) | CT | - | In training set : 0.799 for combined model 0.697 for CT-staging<br><br>In Cross-validation: 0.727 for the combined model 0.628 for CT staging | Combined model better than CT staging only in training set and cross validation set both. |
|  | Huang et al. (10) | CT | - | Discriminative performance for high-grade and low-grade CRAC : 0.735 for training 0.735 for validation<br><br>Distinguishing ability for histologic grade: 0.725 training 0.895 validation | Radiomics could verify high grade and low grade tumor preoperatively. |
|  | Caruso et al. (11) | CT | 107 radiomic features | 0.73 in the internal cohort 0.75 in the external cohort | Radiomics could detect colon high-risk colon cancer. |
|  | Mascarenhas et al. (12) | CCE | - | N/A | AI showed high performance in distinguishing blood from mucosal CRC lesions in CCE. |
|  | Granata et al. (110) | MRI | 851 radiomics features | N/A | T2-W extracted textural images got the best performance. |
|  | Li et al. (111) | CT | 306 CT imaging features | Individual stage classifier (stage IV: [1.00; 1.00]), III: [0.99; 0.99], II: [0.86; 0.99] And multistage classifier Detect KRAS and PNI with the AUC of 0.862 and 0.793 | Individual stage classifier detect better than multistage classifier. |
|  | Huang et al. (112) | conventional CT | - | c-index: 0.803 for validating set and c-index: 0.817 for training | A nomogram based on CEA level radiomics signature showed great performance of risk assessment of PNI in CRC. |
|  | Grosu et al. (113) | CT colonographic | Image texture (n=68) Quantitative image features characterizing shape (n=14) and gray level histogram statistics (n=18), | 0.90 in the size category of 10 mm or larger 0.87 in the size category of 6–9 mm in | Machine learning could differentiate benign and malignant polyps. |

|  | Näppi et al. (19) | CT colonography | - | N/A | CT colonography could detect serrated polyps accurately. |
|---|---|---|---|---|---|
|  | Wesp et al. (114) | CT colonography | - | ROC-AUC of 0.83 for SEG model and ROC-AUC of 0.75 for no SEG | For 90% of polyp tissues. Grad-CAM++heat map score was ≥0.25 showed a good performance of radiomics in automatically detecting high risk polyps. |
|  | Yuan et al. (115) | CT | - | AUC : 0.922 for ResNet3D + SVM classifier model and AUC: 0.791 for routine CT | Admission of SVM classifier+ ResNet3D to a deep leaning algorithm could reliably predict Synchronous Peritoneal Carcinomatosis. |
| **Staging** | Liang et al. (20) | CT | - | AUC of 0.708 for validation dataset AUC of 0.79 staging CRC | Radiomics indicated the great potential to distinguishing stage I-II and stages III-IV CRC patients. |
|  | Castro et al. (21) | CT | - | Kappa was achieved 64.7 % And the accuracy was achieved 83% | Radiomics showed the great performance to detecting gene expressions even without using biopsy or more procedures. |
|  | Wu et al. (22) | DECT | Nine top-ranked features | 0.961 for training 0.918 for validation set | Radiomics based on DECT could detect MSI status in CRC patients accurately. |
|  | Li et al. (116) | contrast-enhanced CT | 1628 features Included Gray Level Co-occurrence Matrix (n = 1518) Gray Level Run Length Matrix (n = 33) shape (n = 18) Intensity Histogram (n = 49) Neighbor Intensity Difference (n = 10) | 0.73 for external validation set 0.79 for training set | A combined radiomics model with clinical features could predict MSI status in CRC patients. |
|  | Fan et al. (24) | CT | 6 radiomics features and 11 clinical features | AUC for combined model: 0.752 Clinical factors: 0.598 and radiomics alone:0.688 | Combination of clinical factors and radiomics can improve the MSI status prediction. |

|  | Pernicka et al. (117) | CT | 254 Intensity-based | AUC of combined model: 0.80 for training set and 0.79 for test set<br><br>Model of radiomics only had an AUC of: 0.76 and clinical only: 0.74 | Study indicated radiomics features and combined model could detect preoperative MSI status more accurate than only clinical factors |
|---|---|---|---|---|---|
|  | Pei et al. (118) | abdominal pelvic computed tomography (CT) | - | 0.74 for training set<br>0.77 of validation set | Nomogram of combination of radiomics and clinical features could improve the prediction of MSI status. |
|  | Hong et al. (109) | CT | - | AUC of combined model: 0.799 for training set 0.697 for CT staging<br><br>Cross validation:<br>0.727 for training set<br>0.628 for CT staging | Radiomics features in combination with CT staging improved the potential of high risk patient identification. |
| Treatment response | Wei et al. (30) | Contrast-enhanced multidetector computed tomography (MDCT) | - | AUC of clinical factor based on CEA level: 0.558 and 0.489 for validation and training set respectively,<br><br>Radiomics model: 0.820 and 0.745<br>Combined method: 0.830 and 0.935 | The DL based model in combination of clinical factors could improve the predicting treatment response of chemotherapy in CRLM patients |
|  | Li et al. (35) | MRI | - | AUCs of the CE-T1WI, DWI,T1WI and, T2WI, in the training set: 0.78, 0.71, 0.74, and, 0.71, respectively<br><br>And in external validation set:<br>0.77 ,0.70 ,0.67, 0.66<br>The AUC of multi-sequences was 0.78 in both training and validation set | Multi-parametric MRI model could improve the prediction of MSI status. |
|  | Giannini et al. (119) | CT | 107 radiomics features | N/A | Delta radiomics could significantly improve the detection of non-response liver metastatic CRC patients to chemotherapy. |
|  | Joa Ahn et al. (38) | CT |  | Validation set<br>0.785 detected for narrower Standard Deviation on 3D<br>0.797 for lower skewness on 2D | CT textural analysis improves the prediction of treatment response cytotoxic chemotherapies procedures. |
|  | R Andersen et al. (39) | CT | - | N/A | Skewness increased during the treatment and spatial scaling factors detected SSF$\geq$1.0. CT textures analyzing could help in prediction of treatment response to the regorafenib |

| Zhang et al. (120) | T2-weighted MRI | 5 histogram features including : kurtosis, mean, skewness, entropy, and variance And 5 gray level co-occurrence matrix features including : GLCM; entropy angular second moment (ASM), , contrast, inverse difference moment (IDM), and correlation | AUC of association of ASM and variance was: 0.814. AUC of 0.602–0.784 was detected for the prediction of good response to chemotherapy for higher variance, IDM , entropy1, lower ASM, contrast, entropy2 and correlation, | The highest AUC for prediction response to chemotherapy was for ASM and variance. |
|---|---|---|---|---|
| Dohan et al. (42) | CT | - | N/A | A radiomic signature including decrease in density, sum of the target liver lesions (STL), and computed textural analysis could predict overall survival and response to chemotherapy |
| Ma et al. (43) | MRI | 396 radiomics features | AUC of Radiomics MRI model achieved to the 0.733 and 0.753 for training and validation set respectively, Radiomics-clinical nomogram detected higher AUC of AUC of 0.809 for responders and non-responders to chemotherapy | The nomogram based on clinical factors CA19 level and radiomics features has a great efficacy for predicting chemotherapy response. |
| Jimenez et al. (44) | T2w MRI | 191 texture descriptors and 198 shape descriptors | 0.73 | Radiomics shape and texture descriptors could find a reliable association between T2W MRI and tumor staging after NCRT. |
| Markich et al. (46) | CT | 64 radiomics features | N/A | Radiomics features before radiofrequency ablation of lung metastasis and after based on early CT follow-up could help in detecting nodules with the high risk of progression. |
| Lu et al. (121) | CT | - | C-index: 0.627 for sized base 0.649 for deep learning network 0.694 for combination of sized based and DL network | The combination of deep learning and clinical factors could predict the response of tumor to the chemotherapy. |
| Dercle et al. (49) | CT | 1757 radiomics features and 1742 pre-trained deep-learning features | Response to anti-epidermal growth factor receptor therapy: $FC^{HQ}$: 0.80 $FC^{SD}$: 0.72  AUC of chemotherapy: $FH^Q$: 0.59 $FS^D$: 0.55 | In cetuximab-using sets, radiomics signatures could predict treatment response and find a great relation to the overall survival. |

| | | | | | |
|---|---|---|---|---|---|
| Metastasis | Gatos et al. (53) | T2-weighted MRI | 42 textural features and 12 morphological features | N/A | Highest accuracies detected for PNN classifier, the overall accuracy was detected 90.1%. it has indicated the powerful performance of radiomics classifications in MRI- based liver status evolutions. |
| | Creasy et al. (122) | CT | 254 image features | N/A | Radiomics tools could predict patients with high risk liver metastasis by assessment of liver parenchyma heterogeneity in image features. |
| | Taghavi et al. (123) | portal venous phase CT | 1767 radiomics features | AUC of clinical features, radiomics features and combination of radiomics and clinical have reported : 71%, 86% and 86% retrospectively | Radiomics assessment of early CT images demonstrated good performance of diagnosing high risk patients for colorectal metastasis. |
| | Alongi et al. (57) | [18F]FDG PET/CT | For per-lesion analysis, only the GLZLM_GLNU feature was selected For PET/CT images three features were selected, For per-patient analysis. Three features for stand-alone PET images one feature for the PET/CT data set. For liver metastasis: in per-lesion analysis one PET feature (GLZLM_GLNU) from PET images and three features from PET/CT data set. And three PET features and a PET/CT feature (HUKurtosis) for liver lesions per-patient analysis | AUROC 65.22% for CT features and PET features AUROC 95.33% For PET/CT features analysis AUROC 88.91% For alone PET imaging | The best performance achieved when using PET/CT features analyzing and showed great performance of prediction of disease progression. |
| | Granata et al. (58) | MRI | 851 radiomics features | N/A | The best accuracy achieved by using KNN and for detection of tumor budding. The radiomics showed great performance in identification of prognostic features of metastatic patients. |
| | Li et al. (124) | CT | - | AUC = 0.855 for clinical-radiomics model AUC = 0.771 for clinical-only model AUC = 0.764 For Radiomics-only model | Radiomic in combination of clinical key factors have reached to the optimal results and have shown great performance of prediction of pretreatment peritoneal metastasis. |

| Study | Modality | Features | Performance | Conclusion |
|---|---|---|---|---|
| Li et al. (60) | CT | 841 | 0.90±0.02 for training set 0.86±0.11 for verification 0.906 for cross-validation set 0.899 For test set | Radiomics nomogram based on clinical risk and radiomics features achieved appropriate results in all datasets. |
| Hu et al. (125) | CT | 203 texture features | 0.929 for training set of clinical-radiomics nomogram and 0.922 for validation cohort | Nomogram based on combination of textures and clinical factors showed great accuracy in discrimination of lung metastasis. |
| Eresen et al. (126) | CT | 146 quantitative CT imaging features | N/A | Radiomics derived model have shown better performance to diagnose the lymph node metastasis compared to the clinical features. |
| Taghavi et al. (127) | CT | 1593 radiomics features | N/A | Between three models including radiomics, clinical model and combined model, combined model has shown the best performance of prediction of tumor progression in liver metastasis patients |
| He et al. (128) | 18F-FDG PET/CT | 70 PET radiomic features | 0.866 for logistic regression 0.903 For XGBoost | Between five machine learning models, XGBoost and logistic regression have shown the best performance in prediction of regional lymph node metastasis |
| Rocca et al. (66) | CT | 22 features | N/A | Formal method could identify colorectal liver metastasis in early stages and heterogeneous clinical samples. |
| Huang et al. (67) | portal venous–phase CT | 24 features | N/A | Nomogram performed appropriately based on the CT-reported LN status, clinical risk factors and radiomics signature |
| Dou et al. (68) | CT | 125 parameters | N/A | Radiomics parameters in combination of clinical parameters could help in T staging of colorectal cancers, and discrimination of high risk and low risk patients. |
| Liu et al.(129) | CT | 1724 radiomics features | N/A | Among these features some of featured stayed stable that could reflect the ley point for distinguishing metastatic nodules from benign. |

| | Rizzetto et al. (130) | contrast-enhanced computed tomography | 32 textural radiomics features including grey level run length matrix (GLRLM) and grey level co-occurrence matrix (GLCM) | N/A | A 2D contouring approach may help mitigate overall inter-reader variability, albeit stable RFs can be extracted from both 3D and 2D segmentations of CRC liver metastases. |
|---|---|---|---|---|---|
| | Bae et al. (131) | contrast-enhanced liver MRI and contrast-enhanced CT | 129 features including: and 10-dimensional shape features and 119-dimensional texture features | 0.9426 | Radiomics model demonstrated great diagnostic performance for verifying benign and metastatic lesions, but it had limitations in differentiating subcentimeter lesions and hemangiomas. |
| **Prognosis and survival** | G. Lubner et al. (132) | CT | - | N/A | Tumor grade of non-treated CRC liver metastasis have an association with the CT textures including MPP, entropy, and SD. |
| | Kang et al. (133) | 18F-FDG-PET | gray-level zone length matrix short-zone low-gray-level emphasis (GLZLM_SZLGE) and gray-level run length matrix long-run emphasis (GLRLM_LRE) | N/A | A radiomics nomogram performed based on nodal stage, lymphovascular invasion, and rad_score demonstrated great prognostic performance in validation set. |
| | Wang et al. (134) | ??? | 8 radiomics features | 0.860 for overall survival and 0.875 for disease free survival | Nomogram based on radiomics, clinical factors, pathomics, and immunoscore could accurately predicting the prognosis of colorectal cancer lung metastasis patients, after procedures. |
| | Mühlberg et al. (135) | contrast-enhanced CT | - | Clinical model: 0.56  whole liver tumor burden (WLTB) alone : 0.53  Radiomics: 0.65  Aerts radiomics prior model: 0.76  And when applied on WLTB: 0.68 tumor burden score (TBS): 0.70 | Geometric metastatic spread (GMS) model could be used at the first step of predicting the risk of CRC patients. |
| | Badic et al. (72) | contrast-enhanced CT | 61 textural features, 11 intensity 1st-order metrics and 16 shape descriptors | AUC of MR was 0.75 And for SVM was 0.64 | Radiomics models based on CT scan could improve the prediction of tumor recurrence after surgery. |

| Author | Modality | Features | Results | Conclusion |
|---|---|---|---|---|
| Huang et al. (73) | CT | 45 radiomics features, including gray-level co-occurrence matrix (GLCM) features and gray-level histogram features | iAUC: 0.841 against refence model 0.763 against 0.785 for the training cohort 0.664 for the external validation cohort 1 0.790 against 0.669 for the external validation cohort 2 | The radiomics model could effectively predict overall survival and utilize in differentiation of CRC tumor stages. |
| Yao et al. (74) | MDCT | - | N/A | Radiomics signature could effectively predict disease free survival of stages II-III colorectal cancer patients. |
| Dai et al. (75) | CT | From 647 screened radiomics 26 relapse-specific features and 13 death-specific | 0.768 for predicting OS and 0.744 for relapse free survival | The specific radiomics signatures including death and relapse features which could significantly improve the prediction ability of CRC prognosis. |
| Cai et al. (76) | CT | 5 radiomics features | 0.734 for primary cohort 0.86 for validation cohort | Radiomics score based on preoperative CT images could play as a prognostic in CRC patients. |
| Chu et al. (77) | CT | 12 features | 0.774 for nomogram of CXCL8-derived and tumor staging | This combination model could reliably predict CRC patient prognosis. |
| Van Helden et al. (136) | [18F]FDG PET/CT | 10 radiomics features | N/A | Research could identify significant association between the tumor spherocity in pretreatment imaging and final improvement of prognosis. |
| Lv et al. (137) | 18F-FDG PET/CT | 1246 | N/A | Pretreatment 18F-FDG PET/CT radiomics could predict prognosis of the CRC patients. |
| Shi et al. (83) | CT | 851 radiomics features | 0.95 for the training cohort 0.79 for validation cohort | Radiomics combined with semantic features can predict gene mutation of CRC patients including KRAS and NRAS and BRAF, that could significantly improve the prediction of survival. |
| He et al. (138) | CT | 1025 features including Texture wavelet characteristics tumor intensity shape and size | ResNet model: 0.90 for testing model 0.93 for admission of ROI and 20-pixel 0.818 for radiomics model | ResNet model have shown better performance in predicting KRAS status mutation in CRC patients, that showed the radiomics ability off prediction gene mutation based on imaging. |

| | | | | | |
|---|---|---|---|---|---|
| | Granata et al. (139) | Contrast Enhanced MRI | 48 texture features | 0.84 for KNN | Radiomics texture analysis in combination with patter recognition methods and adoption of multivariate analysis Could improve the ability of predicting RAS mutation status. |
| | Badic et al. (140) | CT | - | N/A | Gene expression combined with histopathologic factors and radiomics features based on CT have shown a great performance of prediction of CRC patient's prognosis. |
| | Chen et al. (141) | CT | 10 features | 0.788 in internal validation cohort 0.775 in external validation cohort For combined model: 0.777 and 0.767, respectively For clinical model were 0.768 and 0.623 respectively. | The study demonstrated the accuracy of radiomics signature in combination with clinical factors in prediction of MSI status of CRC patients. |
| | Xue et al. (90) | portal-venous CT | 1018 radiomics features | 0.75 for 2D radiomics model 0.84 for 3D radiomics model 0.92 for nomogram model | Radiomics 2D and 3D models could predict KRAS status and clinical-radiomics nomogram demonstrated better predicting performance in prediction of KRAS mutation status. |
| Cancer recurrence | Huang et al. (142) | CT | 1037 radiomic features | 0.56 for CATCH model | Cancer recurrence has been related to ten radiomics textures. That this study has shown the efficacy of adjusting radiomics textures to immune gene expression to prediction of disease recurrence. |
| | Fan et al. (143) | CT | 114 features | For rad-score AUC: 0.886 in training cohort 0.874 in validation cohort For Nomogram 0.954, in training cohort 0.906 in validation cohort | Study identified radiomics in combination of clinical variables could reliably predict the recurrence of colorectal cancer in stage II patients. |
| Neoadjuvant rectal cancer | Song et al. (144) | MRI | 30 radiomics features | Mean AUC : 0.7835 for MRI-based radiomics (MBR) 0.7871 for combination of MBR and clinical baseline characteristics known as (CMBR) 0.7916 in admission of neoadjuvant treatment | Study showed the accuracy and great performance of radiomics based models against clinical parameters in prediction of tumor response to chemotherapy. |
| | Wang et al. (145) | CT | 1130 radiomics features | $0.68 \pm 0.13$ for neoadjuvant rectal score (NAR)>16 model and $0.59 \pm 0.14$ for NAR < 8 model | This study showed the potential of radiomics model in prediction of neoadjuvant therapy response of LARC. |

| Feng et al. (146) | pelvic MRI | 2106 radiomics MRI features | 0.868 in training cohort 0·860 in validation cohort 1 0·872 in validation cohort 2 0·812 in prospective validation | The RAdioPathomics Integrated prediction System called RAPIDS showed significant performance in prediction of pathologic response of LARC to neoadjuvant therapy. |
|---|---|---|---|---|
| Dinapoli et al. (147) | T2-weighted MR scans | 1200 features | 0.73 for internal study 0.75 for external study | The study investigated the power of vendor-independent model in prediction of pathological complete response of LARC to therapies only based on pretreatment MR images. |
| Defeudis et al. (148) | MRI | 157 features | N/A | Radiomics model can help clinicians to improve their decision making and detecting the LARC response to neoadjuvant therapy. |

**Table 1.** The selected studies of radiomics-based artificial intelligence (AI) studies in colorectal cancer.

MRI: magnetic resonance imaging, CT: computed tomography, N/A: not applicable, CRAC: low-grade colorectal adenocarcinoma, CRC: colorectal cancer, CCE: Colon capsule endoscopy, CEA: Carcinoembryonic antigen, PNI: Perineural invasion, SVM: support vector machine, DECT: Dual-energy computed tomography, DWI: diffusion-weighted imaging, PET: Positron emission tomography, T2W: T2-weighted, LN: lymph node, OS: overall survival, DL: deep learning, NCRT: neoadjuvant chemo-radiation therapy, CXCL8: C-X-C Motif Chemokine Ligand 8, LARC: long-acting reversible contraception, 18-FDG: Fludeoxyglucose F18, MSN: Microsatellite instability, KNN: K-Nearest Neighbors, KRAS: Kirsten rat sarcoma viral oncogene homolog, AUC: area under the curve, CATCH: Classification in the high-dimension, CRLM: colorectal lymph node metastasis, PNN: Probabilistic Neural Network